\documentclass{aa}  
\usepackage{graphicx}
\usepackage{txfonts}
\usepackage{hyperref}
\usepackage{natbib}
\usepackage{xcolor}
\usepackage{textgreek}
\usepackage{lipsum}
\usepackage{verbatim}

 \def\simle{\mathrel{\hbox{\rlap{\hbox{\lower4pt\hbox{$\sim$}}}\hbox{$<$}}}}
 \def\simgr{\mathrel{\hbox{\rlap{\hbox{\lower4pt\hbox{$\sim$}}}\hbox{$>$}}}}

\newcommand{\msol}{M_\odot}

\newcommand{\days}{\,\mathrm{d}}

\begin{document} 

\title{Analytic approximations for massive close post-mass transfer binary systems}

\author{C.~Schürmann\inst{1}\fnmsep\thanks{email: \texttt{chr-schuermann@uni-bonn.de}}
\and N.~Langer\inst{1}\fnmsep\inst{2}
\and J.~A.~Kramer\inst{2}
\and P.~Marchant\inst{3}
\and C.~Wang\inst{4}
\and K.~Sen\inst{5}
}

\institute{Argelander-Institut für Astronomie, Universität Bonn, Auf dem Hügel 71, 53121 Bonn, Germany
\and Max-Planck-Institut für Radioastronomie, Auf dem Hügel 69, 53121 Bonn, Germany
\and Instituut voor Sterrenkunde, KU Leuven, Celestijnenlaan 200D, 3001 Leuven, Belgium
\and Max Planck-Institut für Astrophysik, Karl-Schwarzschild-Straße 1, 85748 Garching, Germany
\and Institute of Astronomy, Nicolaus Copernicus University, Grudziadzka 5, 87-100 Toru\'n, Poland
}

\authorrunning{C. Schürmann et al.}
\date{Submitted ??? / Accepted ???}

\abstract{Massive binary evolution models are needed to predict massive star populations in star forming galaxies, the supernova diversity, and the number and properties of gravitational wave sources. Such models are often computed using so called rapid binary evolution codes, which approximate the evolution of the binary components based on detailed single star models. However, about one third of the interacting massive binary stars undergo mass transfer during core hydrogen burning (Case~A mass transfer), whose outcome is difficult to derive from single star models. Here, we use a large grid of detailed binary evolution models for primaries in the initial mass range 10 to $40\msol$ of LMC and SMC composition, to derive analytic fits for the key quantities needed in rapid binary evolution codes, i.e., the duration of core hydrogen burning, and the resulting donor star mass. Systems with shorter orbital periods produce up to 50\% lighter stripped donors and have a up to 30\% larger lifetime than wider systems. We find that both quantities depend strongly on the initial binary orbital period, but that the initial mass ratio and the mass transfer efficiency of the binary have little impact on the outcome. Our results are easily parameterisable and can be used to capture the effects of Case~A mass transfer more accurately in rapid binary evolution codes.}


\maketitle

\section{Introduction}

Massive stars are key constituents of the Universe, as they produce heavy elements, drive the cosmic matter cycle in galaxies, and are the origin of supernovae, black holes, and other spectacular phenomena \citep[e.g.][]{2012ARA&A..50..107L}. It has become clear that most massive stars are born in binary or multiple systems \citep{1998NewA....3..443V,2012Sci...337..444S,2017ApJS..230...15M,2022A&A...658A..69B}. Since stars tend to increase their radius during their life, most binary stars are expected to interact sooner or later, drastically altering the course of their evolution \citep{1992ApJ...391..246P,2013ApJ...764..166D}.

Stellar evolution codes have been constructed, which are capable of predicting the progression of the properties of both stellar components and of the binary orbit in detail --- even though using various physical approximations \citep{1994A&A...290..129V,2001ApJ...552..664N,2001A&A...369..939W,2008MNRAS.384.1109E}. This includes the mass transfer phases, as long as mass transfer does not become dynamically unstable. In particular the numerically robust MESA code, which can compute highly resolved models of both stars in a binary simultaneously, has been extended to include a large spectrum of binary physics \citep{2011ApJS..192....3P,2013ApJS..208....4P,2015ApJS..220...15P}. 

In order to derive population synthesis predictions, several of these codes have been used to produce large grids of binary evolution models \citep{1997A&A...317..487V,2007A&A...467.1181D,2017PASA...34...58E,2020A&A...638A..39L,2020ApJ...888L..12W,2023ApJS..264...45F}. Because their initial parameter space is so much larger than that of single star models, comprehensive grids sufficiently dense to produce well resolved population predictions need $10^4$ to $10^5$ individual detailed binary evolution models \citep[e.g.][]{2020A&A...638A..39L}, constituting a considerable effort. While these efforts have been successful in providing new and important predictions \citep[e.g.][]{2020ApJ...888L..12W,2022A&A...659A..98S}, they are hampered by our assumptions on unconstrained essential physics parameters, for single star and binary evolution physics. It is currently still prohibitively time consuming to perform the required parameter studies with large detailed binary model grids. 

For this reason, so called rapid binary evolution codes have been developed. In most of these, a star is just resolved by two grid points, representing the stellar core and the envelope, and their properties as function of time are approximated from single star models, either analytically or interpolated from detailed single star models \citep[e.g.][]{1996A&A...310..489L,hurley00,hurley02,2006A&A...460..565I,2017NatCo...814906S,2018MNRAS.481.4009V,2018MNRAS.481.1908K,2021ApJ...908...67S,2022ApJS..258...34R,2023MNRAS.524..245R}. While this can not describe the short term thermally unstable evolutionary phases of stars, including phases of mass transfer, it may capture the essential result of mass transfer well enough in most cases, i.e., when the mass donor is essentially stripped of its complete envelope. 

However, mass transfer during core hydrogen burning \citep[Case~A mass transfer, e.g.][]{1994A&A...290..119P} is particularly unruly, since a clear division of the donor star into a core and an envelope is only possible after core hydrogen exhaustion. While Case~A mass transfer occurs only in rather short period binaries, those are favoured by the initial orbital period distribution, such that it concerns about one third of all interacting massive binary stars \citep{2012Sci...337..444S,2013A&A...550A.107S,2014ApJ...782....7D}, or even the majority above about $40\msol$ \citep{2023A&A...672A.198S}. While many rapid codes treat Case~A mass transfer as if core hydrogen burning was already over at the onset of mass transfer, we show below that this can lead to large errors in the predicted donor masses and ages after the mass transfer. In particular, the post-mass transfer donor properties in Case~A binaries are known to strongly depend on the initial orbital period of the binary \citep[cf., fig.~14 of][]{2001A&A...369..939W} and can not be easily derived from single star models. This has important implications for the final fate of the donor stars, as one can see in fig.~B.1 of \citet{2020A&A...638A..39L}, where Case~A models produce neutron stars and Case~B (mass transfer after central hydrogen exhaustion) models produce black holes. This directly effects the predicted number of black holes and neutron stars.

To remedy this problem, we make use of existing large binary evolution model grids computed with MESA, to derive analytic predictions for the key quantities of donor stars directly after Case~A mass transfer, as function of the initial binary parameters. We briefly discuss the key physics and initial parameters of these grids in Sect.~\ref{sec:method}. In Sect.~\ref{sec:results}, we explore the dependencies of the donor properties on the initial binary parameters, and derive analytic fits to our main results. We discuss caveats and uncertainties in Sect.~\ref{sec:discuss}, before we give our conclusions in Sect.~\ref{sec:concl}.

In this paper, we neither investigate the properties of the accretor, as they were covered by e.g. \citet{2021ApJ...923..277R,2023ApJ...942L..32R}, nor contact binaries, as contact alters the course of evolution and they are expected to merge sooner or later \citep{2021MNRAS.507.5013M}.

\section{Detailed binary model grids}\label{sec:method}

We use the grids of detailed binary models calculated by \citet[][see also \citet{2020A&A...638A..39L} and \citet{2022A&A...659A..98S}]{2017PhDT.......434M} with Large Magellanic Cloud (LMC) metallicity and by \citet{2020ApJ...888L..12W} with Small Magellanic Cloud (SMC) metallicity, using MESA version 8845 \citep{2011ApJS..192....3P,2013ApJS..208....4P,2015ApJS..220...15P}. The LMC grid contains models with initial primary (i.e. the initially heavier component of the binary) masses from $10\msol$ to $40\msol$ with initial orbital periods from $10^{0.15}\days=1.4\days$ to $10^{3.5}\days=3162\days$ and initial mass ratios (mass of the initially less massive star over the mass of the primary) from 0.25 to 0.975. The SMC grid contains initial primary masses from $5\msol$ to $100\msol$ with initial orbital periods from $1\days$ to $10^{3.5}\days=3162\days$ and mass ratios from 0.3 to 0.95. We use all models of these grids which undergo Case~A mass transfer with donor masses between $10\msol$ to $40\msol$, as the models outside of this range tend not to yield a stripped donor star, either due to physical (no stable mass transfer) or numerical reasons, see \citet{2017PhDT.......434M}. An extension of the LMC grid by \citet{2022A&A...667A..58P} will be used in Sect.~\ref{sec:fits} to test if our results are applicable outside the adopted mass range. The upper initial period limit for Case~A is a function of donor mass as discussed in Sect.~\ref{sec:fits}.

The initial chemical composition of the models is as in \citet{2011A&A...530A.115B}, and custom-built OPAL opacities \citep{1996ApJ...464..943I} were used to match the initial abundances. The models were computed using the standard mixing-length theory with $\alpha_\mathrm{ml}=1.5$, the Ledoux criterion for convection and step-overshooting with $\alpha_\mathrm{ov}=0.335$ \citep{2011A&A...530A.115B}. We assume thermohaline mixing following \citet{2010A&A...521A...9C} with $\alpha_\mathrm{th}=1$ and apply semiconvection with $\alpha_\mathrm{sc}=0.01$ for the LMC \citep{1991A&A...252..669L} and $\alpha_\mathrm{sc}=1$ for the SMC \citep{1983A&A...126..207L}. The effect of the difference in semiconvection is small during hydrogen burning in the considered donor models \citep{2019A&A...625A.132S}.

The initial spin of both stars is assumed to be synchronous with the orbit \citep{2020A&A...638A..39L} and the tides are treated as in \citet{2008A&A...484..831D}. Differential rotation, rotational mixing \citep[with the ratio of the ratio of the turbulent viscosity to the di†usion coefficient $f_c=1/30$,][]{1992A&A...253..173C} and angular momentum transport are modelled as in \citet{2000ApJ...528..368H,2005ApJ...626..350H} including the Taylor-Spruit dynamo \citep{2002A&A...381..923S}. During Roche-lobe overflow (RLO), the secondary star accretes matter either ballistically or from a Keplerian disk \citep{2005A&A...435.1013P} based on the results from \citet{1975ApJ...198..383L} and \citet{1976ApJ...206..509U}. Rotationally enhanced mass loss \citep{1998A&A...329..551L} stops accretion when the accretor reaches critical rotation \citep{2012ARA&A..50..107L}. The material that has not been accreted leaves the system with the specific orbital angular momentum of the accretor following \citet{1997A&A...327..620S}. If the combined luminosity of both stars does not provide enough energy to unbind the excess material from the system, the calculations were stopped \citep[see eq.~2 of][in particular]{2022A&A...659A..98S}. Models in which overflow at the outer Lagrange point or reverse mass transfer occurs where terminated, too. The remaining models were calculated at least up to central helium depletion.


\section{Results}\label{sec:results}


Case~A mass transfer is rather complex, in that it is composed of three distinct phases \citep{1994A&A...290..119P,2001A&A...369..939W}. It starts with a phase of rapid mass transfer (fast Case~A), which proceeds on the Kelvin-Helmholtz timescale of the mass donor, during which the donor is stripped of a large fraction of its envelope mass. For shorter initial orbital periods, this rapid mass transfer happens earlier during the core hydrogen burning evolution of the donor. It is followed by a nuclear timescale mass transfer phase, driven by the slow expansion of the donor star (slow Case~A), which ends due to its overall contraction of the donor near core hydrogen exhaustion. Immediately thereafter, another rapid mass transfer occurs, driven by the expansion of the remaining hydrogen-rich envelope due to the ignition of shell hydrogen burning. 

This third mass transfer episode (often called Case~AB), which concludes Case~A mass transfer, strips the donor star so much that its envelope mass becomes very small, and it can be approximated for many purposes as a helium star \citep[see however][]{2020A&A...637A...6L,2021A&A...656A..58L}. This is analogous to the situation after Case~B mass transfer, which occurs in binaries which have sufficiently large orbital periods that the donor star avoids mass transfer during core hydrogen burning. However, while in Case~B systems the mass of the stripped helium star closely follows the helium core mass--initial mass relation of single stars, the helium stars emerging from Case~A binaries do not obey this relation. 
Similarly, the age of a donor star at the end of the Case~B mass transfer is very close to the core hydrogen burning life time of a single star of the same initial mass. However, since Case~A donors undergo part of their core hydrogen burning with a significantly reduced mass, their ages at the end of Case~A mass transfer are larger than those of corresponding single stars. Both effects are shown in detail in the following.

\subsection{Analysis of the MESA models}\label{sec:analysis}

\begin{figure*}[h!]
    \includegraphics[width=0.5\hsize]{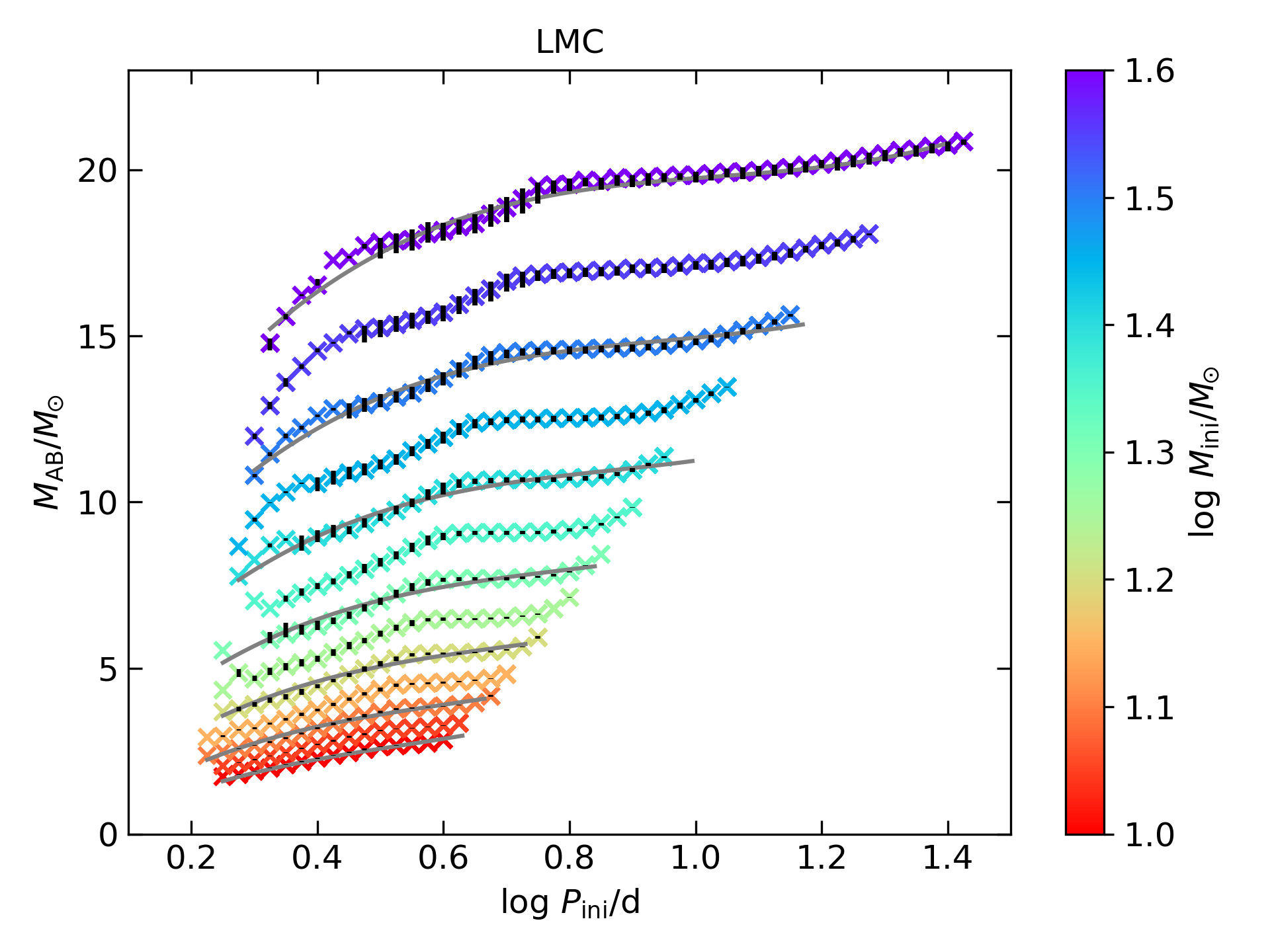}
    \includegraphics[width=0.5\hsize]{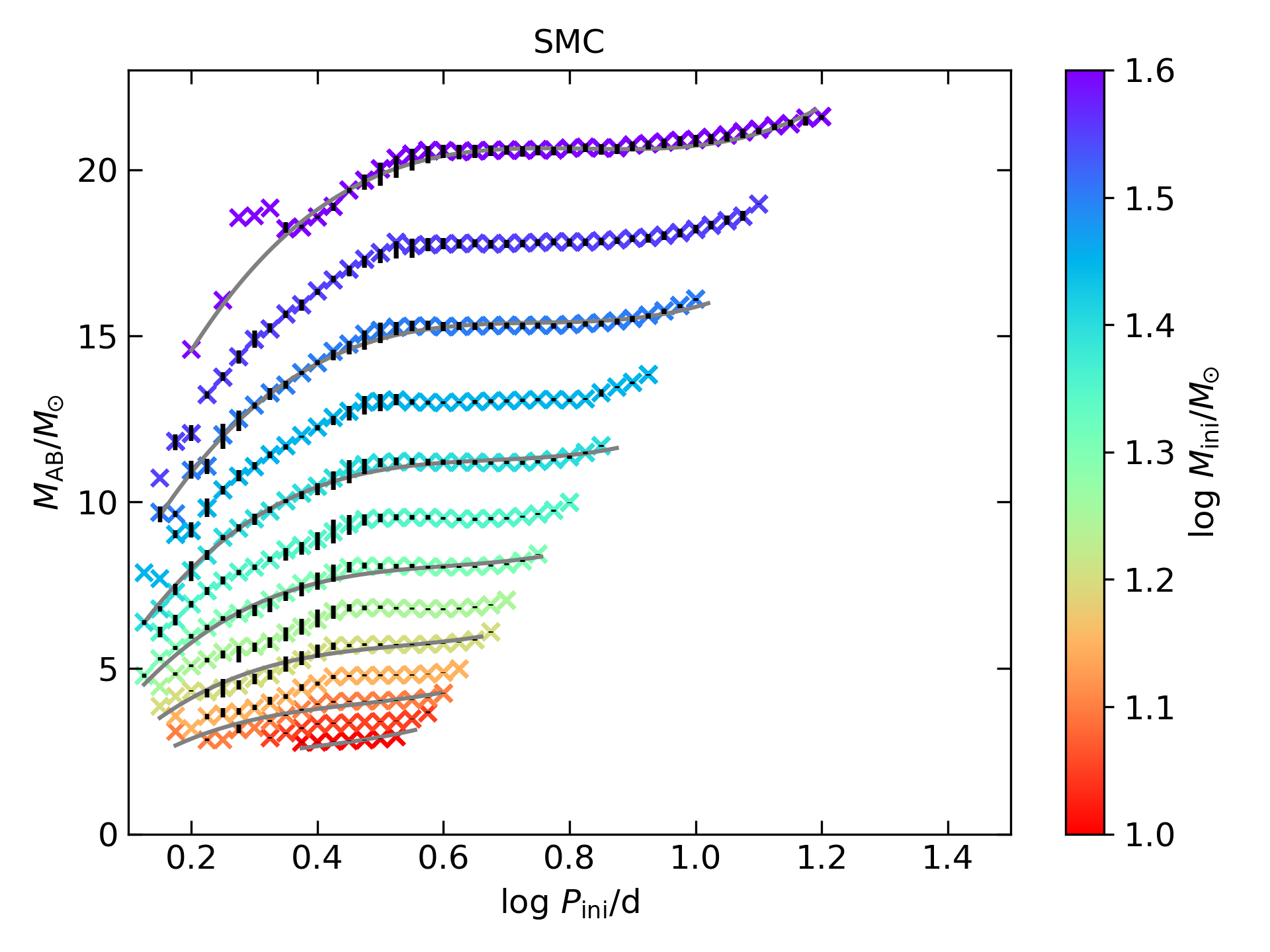}
    \includegraphics[width=0.5\hsize]{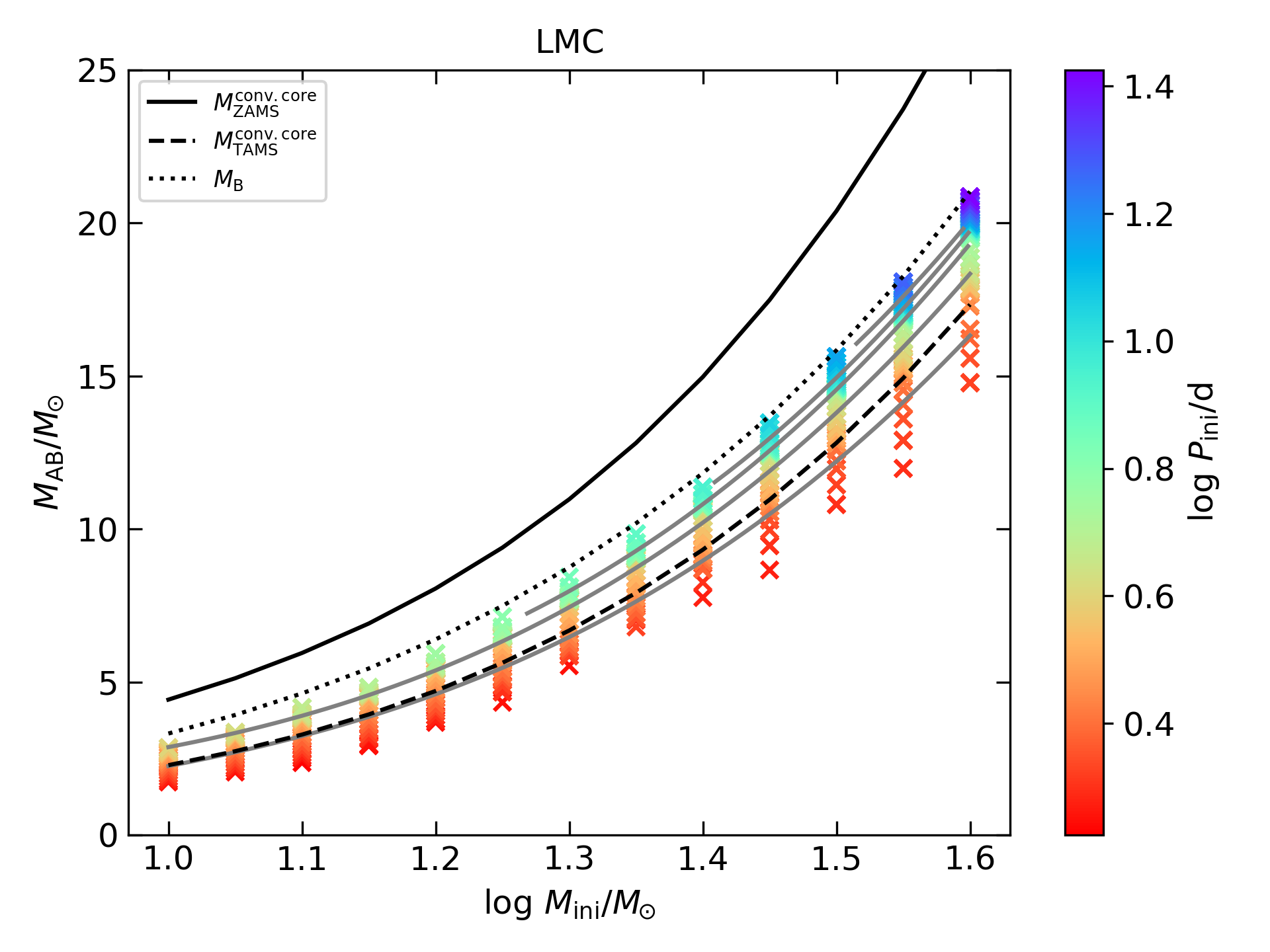}
    \includegraphics[width=0.5\hsize]{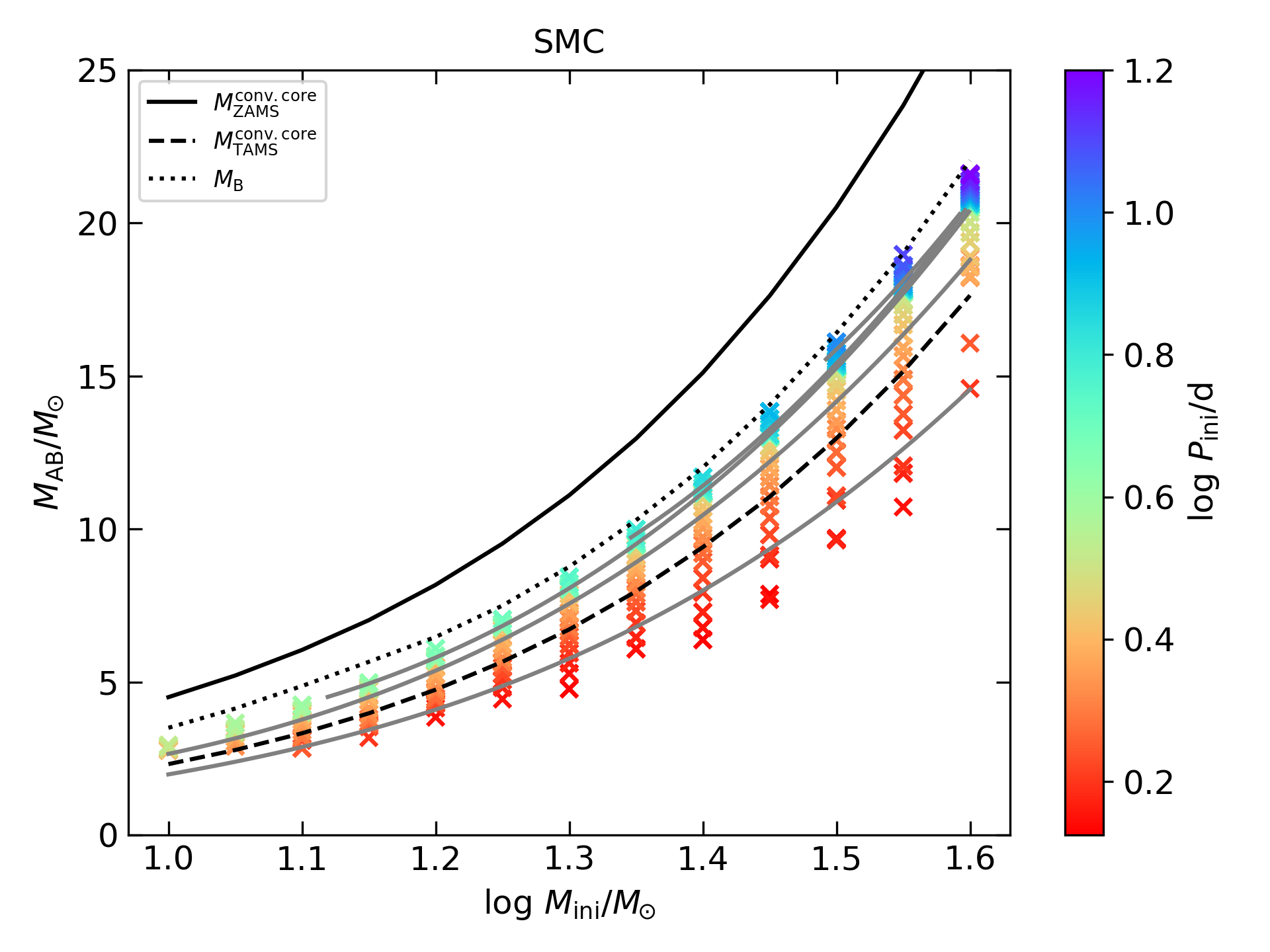}
    \includegraphics[width=0.5\hsize]{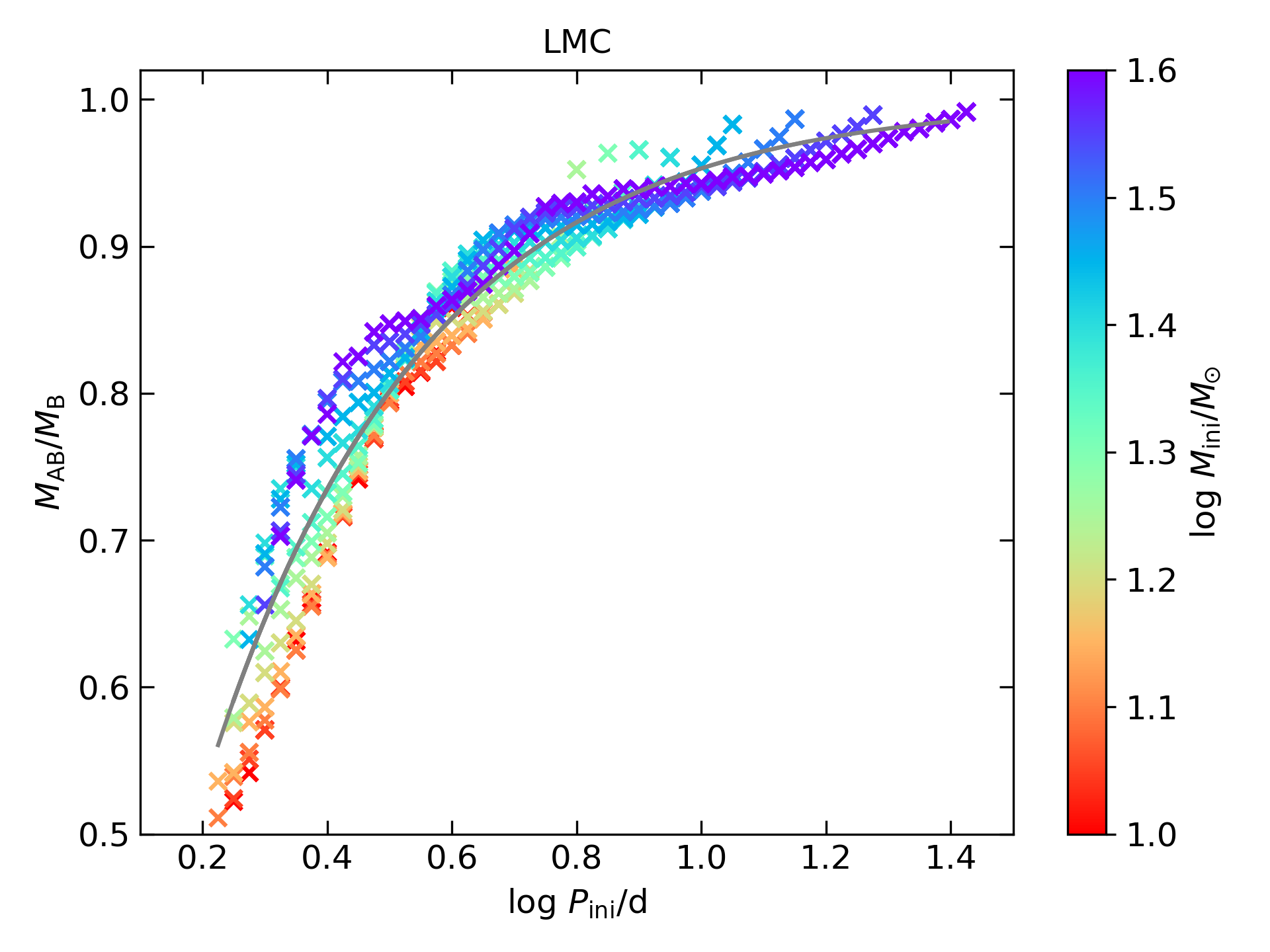}
    \includegraphics[width=0.5\hsize]{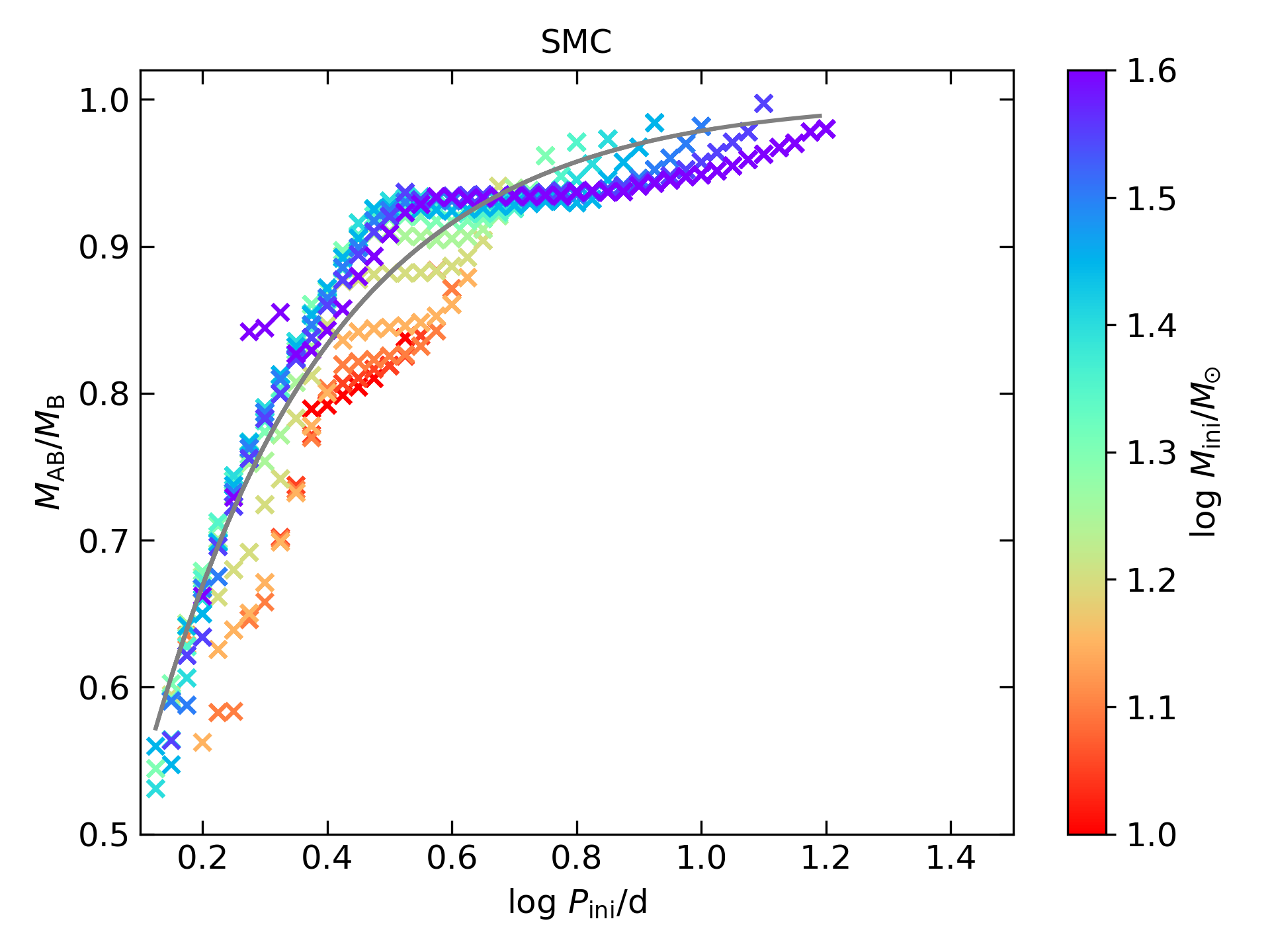}
    \caption{Donor mass immediately after Case~AB mass transfer ($M_\mathrm{AB}$) in units of the Solar mass (top and middle) and in units of the donor mass after Case~B mass transfer ($M_\mathrm{B}$, bottom), as function of the initial orbital period $P_\mathrm{ini}$, with the initial donor mass $M_\mathrm{ini}$ colour coded (top and bottom). The middle panel shows $M_\mathrm{AB}$ as functions of the initial donor mass, where models with the same initial orbital period are indicated with the same colour. Each cross represents the median of $M_\mathrm{AB}$ across different initial mass ratios and in the top plots we indicted in black its interquartile range (distance from first to third quartile of the mass ratio distribution). In the middle plots the black lines show the mass of the convective cores of single stars at the beginning ($M_\mathrm{ZAMS}^\mathrm{conv.core}$) and the end ($M_\mathrm{TAMS}^\mathrm{conv.core}$) of core hydrogen burning, as well as the mass after Case~B mass transfer, as function of the initial stellar mass. Grey lines indicate our best fit to the data. The panels on the left show LMC models, and on the right SMC show models.}
    \label{fig:dm}
\end{figure*}

For this analysis, we define the beginning of a Case~A RLO as the moment where the donor fills more than 99.9\% of its Roche lobe during central hydrogen burning. As the end of Case~AB we use the time when the donor star becomes smaller than 99\% of its Roche lobe after central helium ignition (central carbon abundance surpasses 0.1\%). We found that these assumptions ensured the best tracking of the RLO in our models.

Fig.~\ref{fig:dm} shows the post-Case~AB donor masses $M_\mathrm{AB}$ of all donors in the considered binary model grids, for LMC and SMC metallicities, as function of their initial mass $M_\mathrm{ini}$ and initial orbital period $P_\mathrm{ini}$. In this figure, we have depicted the median values of the post-Case~AB masses across different mass ratios for binaries with the same initial donor mass and orbital period, to enhance the clarity. One can see from the top panels, where we display the interquartile range (i.e. first to third quartile of the mass ratio distribution), that the scatter in post-Case~AB donor masses (for a fixed initial donor mass and initial orbital period) from different initial mass ratios is very limited. Around an orbital period of roughly $10^{0.5}\days\approx3\days$ the interquartile range is for both metallicities slightly larger than elsewhere.
See Sect.~\ref{sec:qeps} for a further discussion.

The top and middle panels of the figure show that, as expected, the post-Case~AB donor masses depend strongly on the initial donor mass. However, on top of that, a clear dependence on the initial orbital period can also be seen. The latter effect is largest for the largest initial donor mass ($\sim 40\msol$), for which the LMC post-Case~AB donor masses cover the range from $14.8\msol$ to $20.9\msol$. For $10\msol$ donors, the post-Case~AB donor masses are found to range from $1.7\msol$ to $2.8\msol$, such that the relative variation is as large as it is for the $40\msol$ donors. For SMC metallicity we find slightly different masses. The post-Case~AB masses of the $40\msol$ donors are $14.6\msol$ to $21.8\msol$, and for the $10\msol$ donors only $2.8\msol$ to $2.9\msol$, due to a smaller number of models surviving the RLO. For all models a small hydrogen-rich layer remains on the donor. 
In the middle panels, we also indicate the convective core mass at be beginning and the end of core hydrogen burning for single stars of the same initial mass. For a given initial donor mass, the largest post-Case~AB mass is always clearly smaller than the initial convective core mass and the convective core mass at central hydrogen exhaustion is only loosely related to the smallest post-Case~AB mass, since central hydrogen exhaustion in single star evolution and Case~AB evolution have followed different evolutionary paths. The donor mass after Case~B mass transfer with same initial donor mass is a much better indicator for the behaviour of the post-Case~AB mass. It is either equal (upper end of the initial donor mass range) or slightly smaller (lower end) than the largest post-Case~AB mass at same initial donor mass. For the SMC models this difference between post-Case~B mass and largest post-Case~AB mass is larger.

Inspired by that, in the bottom panels of Fig.~\ref{fig:dm}, we have scaled the post-Case~AB donor mass to the post-Case~B mass of a model with same initial mass. Interestingly, this ratio shows a very high (but non-linear) correlation with the initial orbital period. This behaviour is more pronounce for the LMC models than for the SMC models. The larger scatter for the SMC models may arise from the post-Case~B mass of the lighter models being heavier than the heaviest post-Case~AB models of the same initial mass. This causes those models to deviate from the curve. Towards smaller initial orbital periods, the ratio of post-Case~AB to post-Case~B mass decreases and, as expected, the ratio converges towards unity for large orbital period. The main difference between the two metallicities is that Case~A occurs at slightly lower initial orbital periods for the lower metallicity. This shift causes the post-Case~AB mass to be higher for the lower metallicity at the same orbital periods. The underlying reason is that for the same initial orbital period the SMC donor fills its Roche lobe later into central hydrogen burning than a corresponding LMC donor, because SMC models are more compact. Thus the SMC donor resembles to a LMC donor at higher initial orbital period.


\begin{figure}
    \centering
    \includegraphics[width=\hsize]{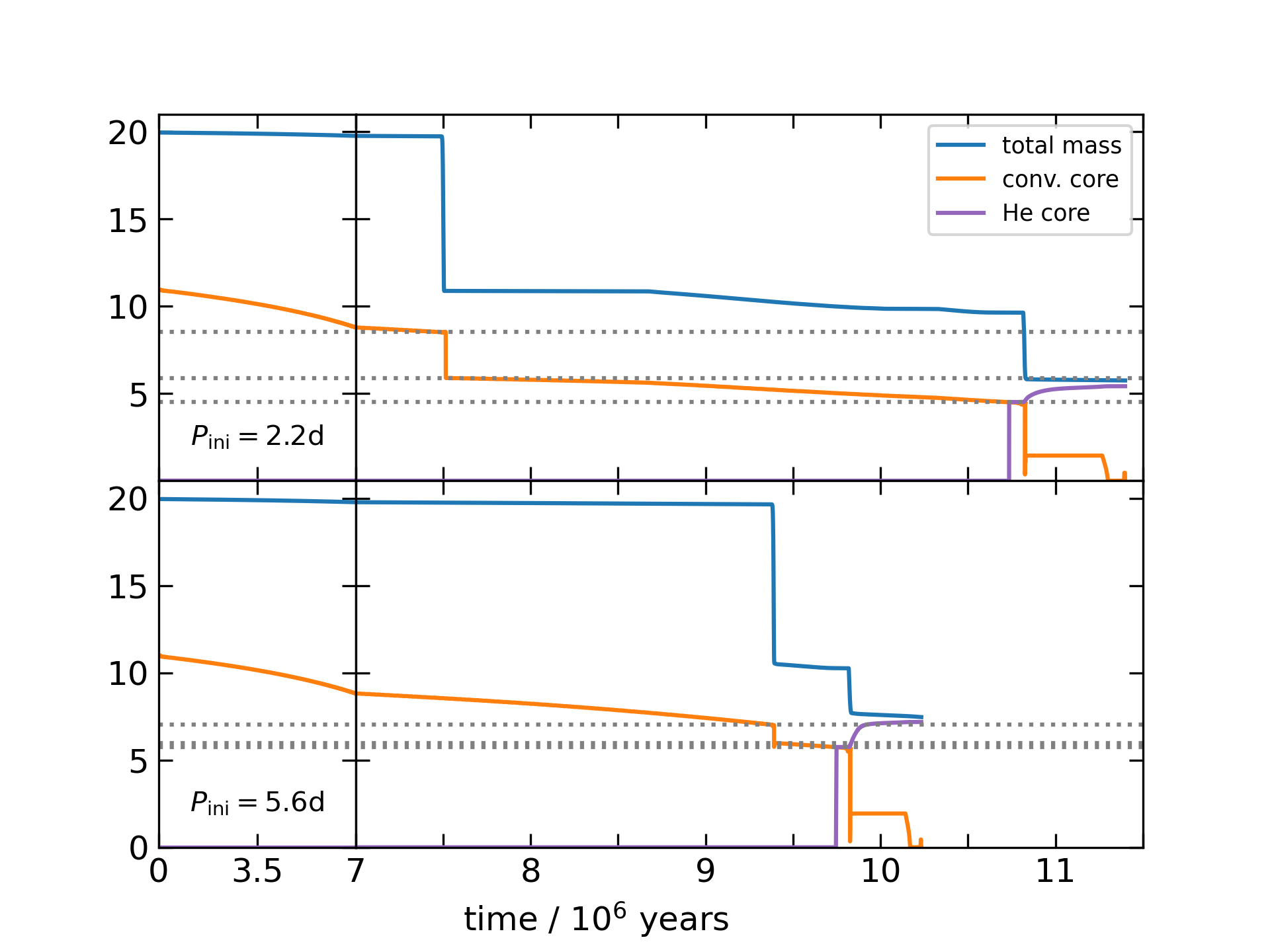}
    \caption{Evolution of the total mass, the convective core mass, and the helium core of the donor model with an initial mass of $20\msol$ with a companion of initially $14\msol$ and an initial orbital period of $10^{0.35}\days=2.2\days$ (top panels) and $10^{0.75}\days=5.6\days$ (bottom panels). We indicate the mass of the convective core at the onset of mass transfer, just after the fast Case~A, and at central hydrogen depletion by grey lines.}
    \label{fig:kipp}
\end{figure}

The period dependence of the post-Case~AB donor masses can be understood as follows. For a given initial donor mass both the mass of the initial convective core\footnote{This includes the overshooting region above and not just the MESA output \texttt{mass\_conv\_core}.} and the donor mass after fast Case~A mass transfer barely depend on the initial orbital period \citep[fig.~F.3 of][]{2022A&A...659A..98S}. This can be seen in Fig.~\ref{fig:kipp}, where we find for an initial donor mass of $20\msol$ a donor mass after fast Case~A of $10.9\msol$ for a small initial orbital period (top panels) and for a wider Case~A system (bottom panels) we get $10.5\msol$. In both models the initial convective core mass was $11.0\msol$ and the convective core masses just before the onset of mass transfer were $8.5\msol$ and $7.0\msol$, as expected since the extend of the convective core shrinks during main-sequence evolution and the mass transfer happens later during hydrogen burning for the wider system. During the fast Case~A phase the mass of the convective core decreases abruptly, namely by $2.6\msol$ for the close system and by $1.1\msol$ for the wide system. The extent of the abrupt shrinking (in mass) depends on how early the Case~A mass transfer occurs i.e. on the initial orbital period of the binary. The shorter the initial orbital period, the greater is the shrinking in terms of mass. This period dependent jump in the convective core mass is the first reason for the period dependence in the post-Case~AB mass. Over the whole model set, it takes values from $0.8\msol$ to $2.6\msol$

Since the remaining central hydrogen burning time is larger for the donor in the closer system ($3.3$ but only $0.4 \cdot 10^6$ years for the wide system), the mass of the donor's convective core decreases even more. During the slow Case~A phase, the donor transfers mass on a nuclear timescale, wherefore the donor in the close system loses more mass. In the example in Fig.~\ref{fig:kipp}, the donor mass at central hydrogen depletion is $9.6\msol$ for the close and $10.3\msol$ for the wide system. This causes the mass of the convective core to become even smaller, which forms the second reason for the period dependency. At central hydrogen depletion, the convective cores has shrunken by $1.4\msol$ since end of fast Case~A for the close system and by only $0.2\msol$ for the wide system. Over the whole model set, this effect can shrink the convective core up to $5\msol$ for the closest and heaviest systems. For light donors, both effect are equally important, since for close and wide systems the difference in mass change of the convective core for the first effect is about $1\msol$ as it is for the second one, while for heavier donors, the second one dominates. Finally, that mass of the convective core at hydrogen exhaustion determines the mass of the helium core, which then determines the mass of the donor after Case~AB mass transfer.

\begin{figure*}[h!]
    \includegraphics[width=0.5\hsize]{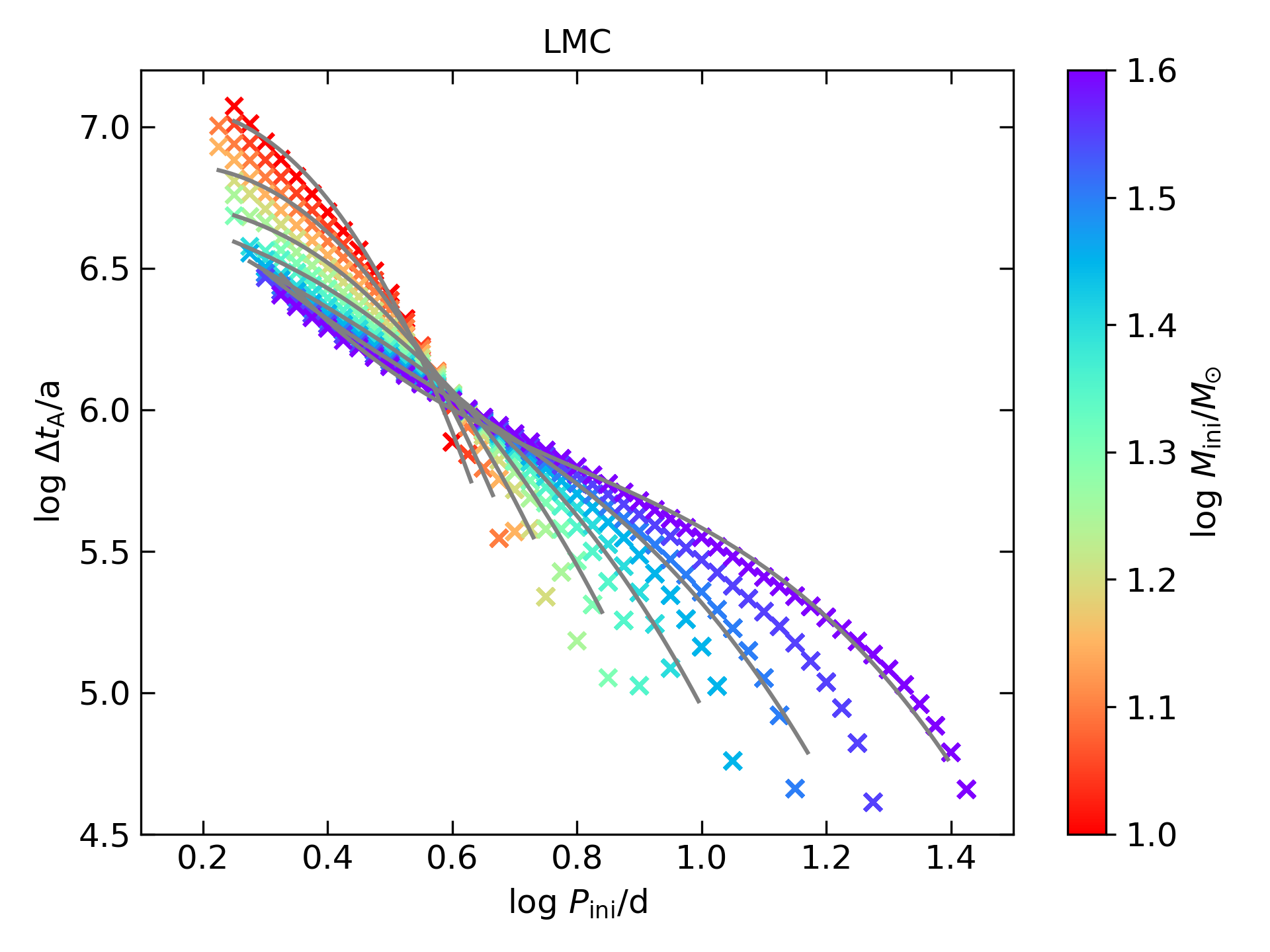}
    \includegraphics[width=0.5\hsize]{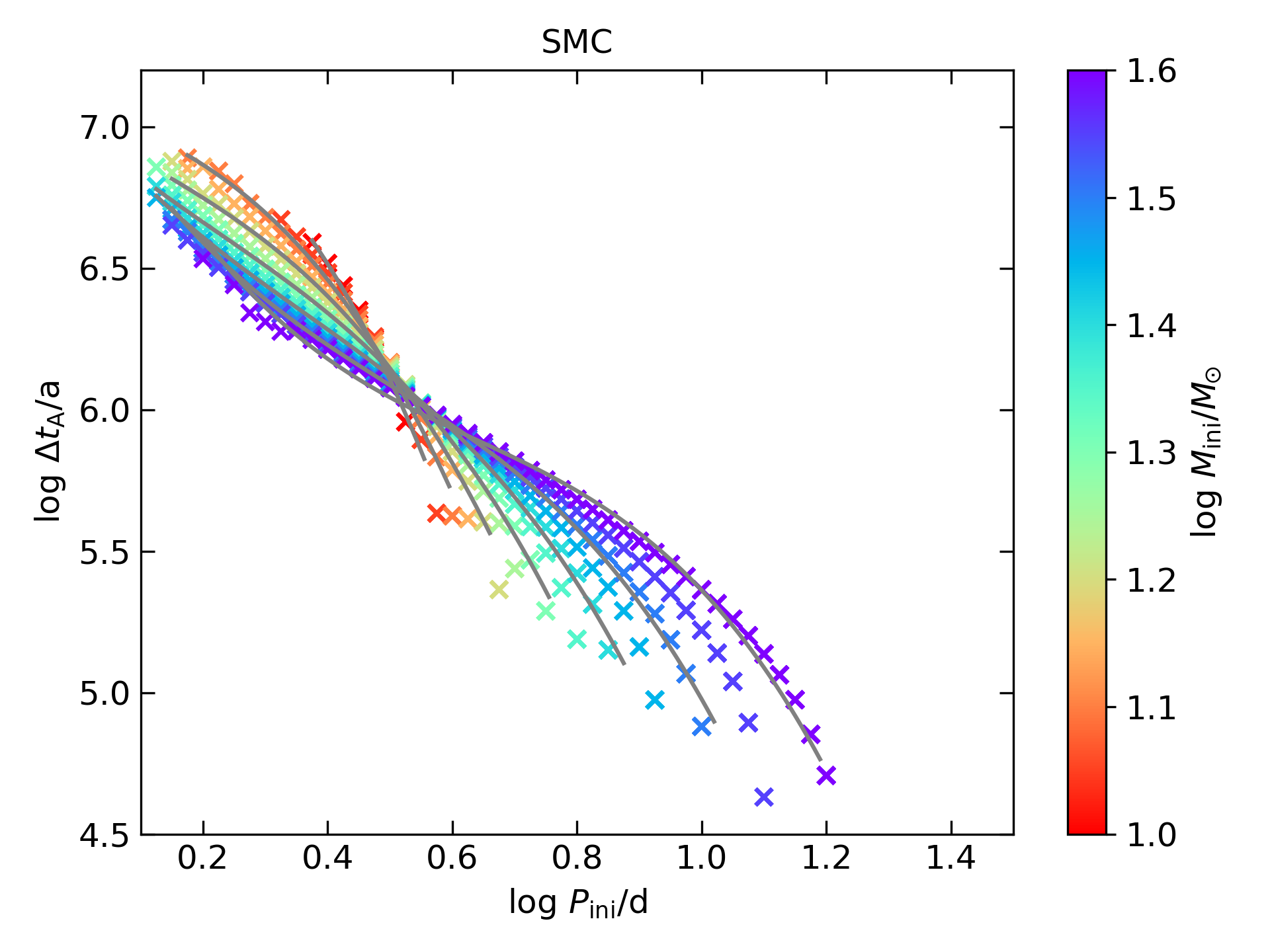}
    \includegraphics[width=0.5\hsize]{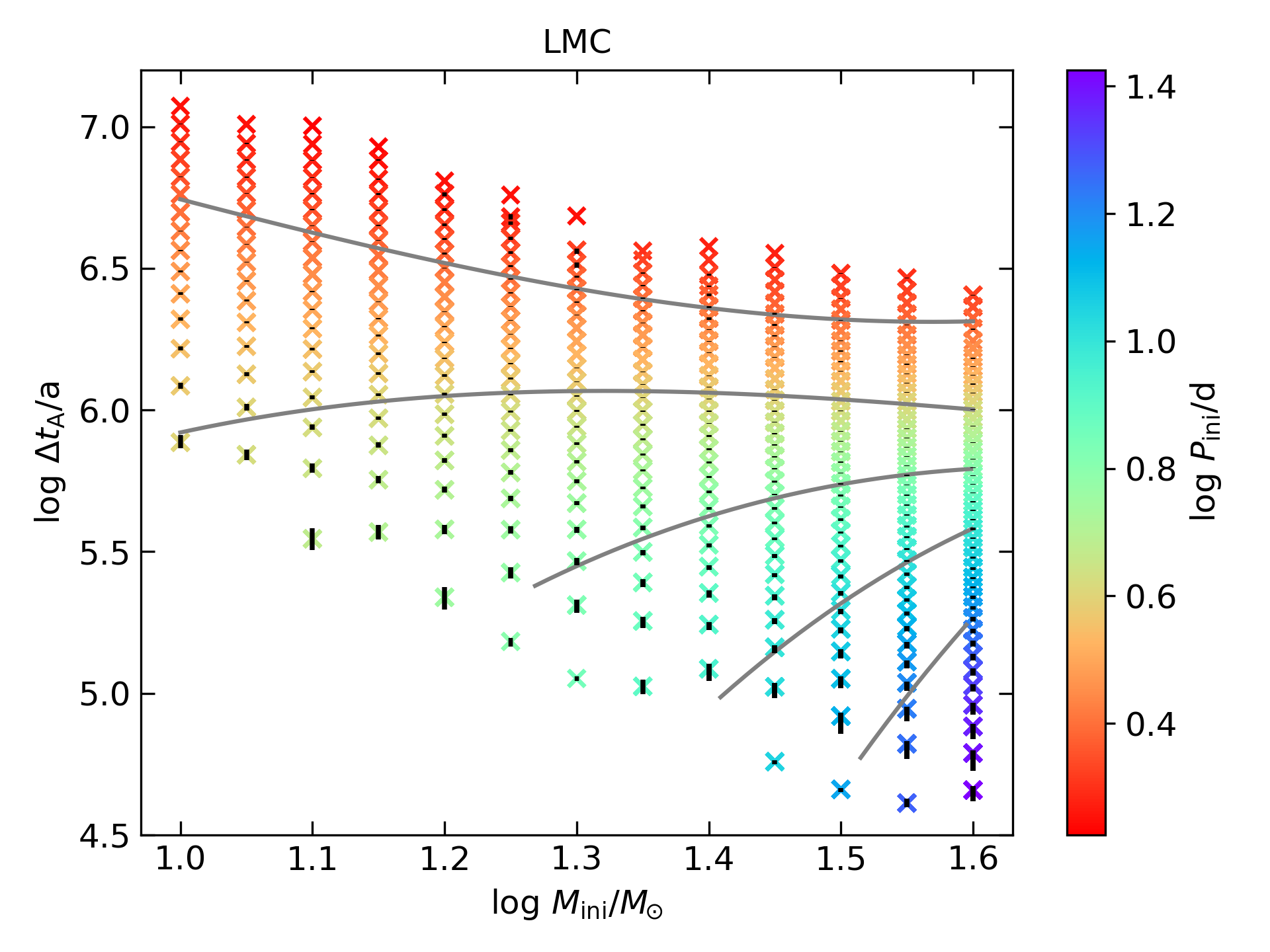}
    \includegraphics[width=0.5\hsize]{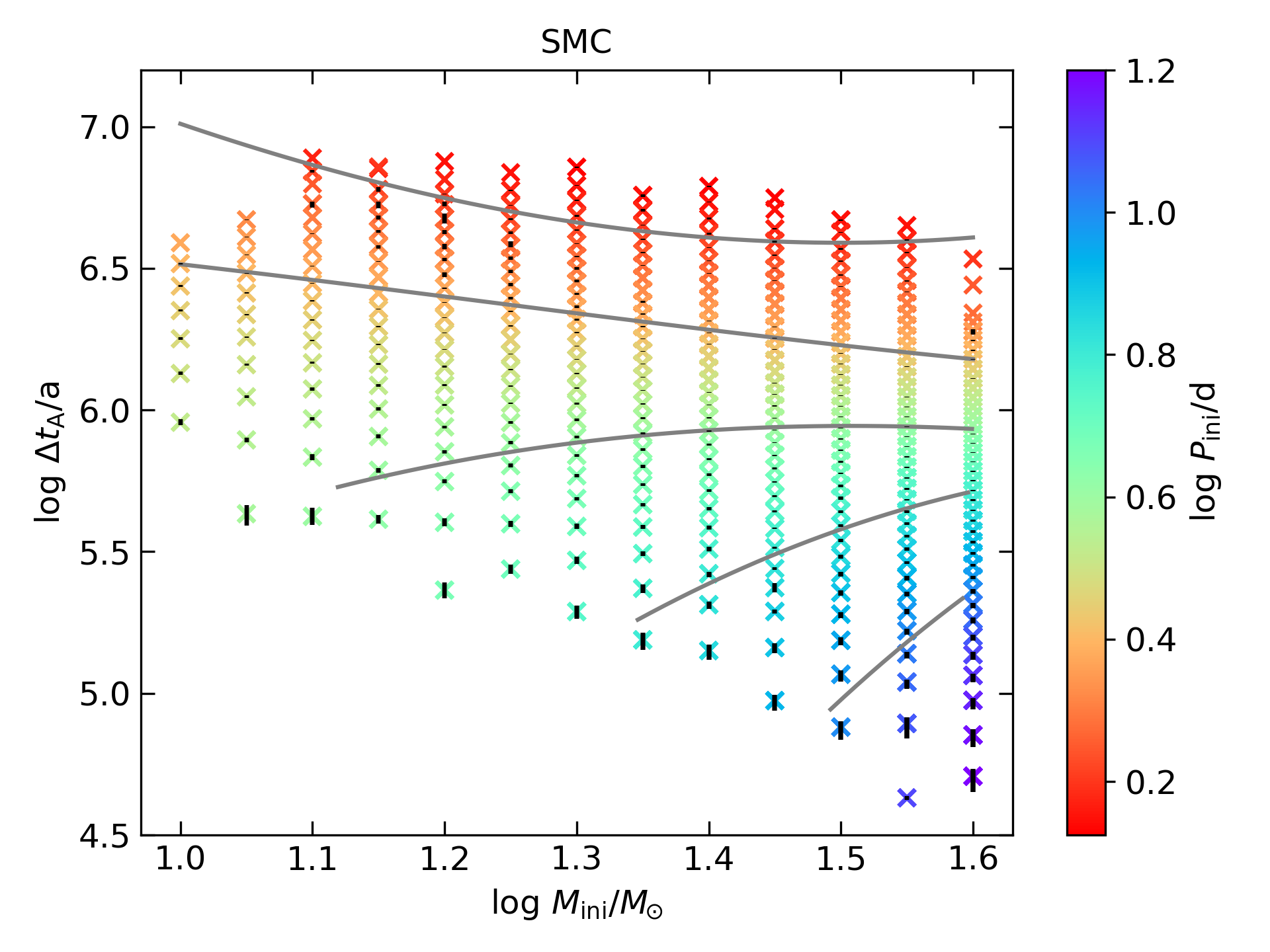}
    \includegraphics[width=0.5\hsize]{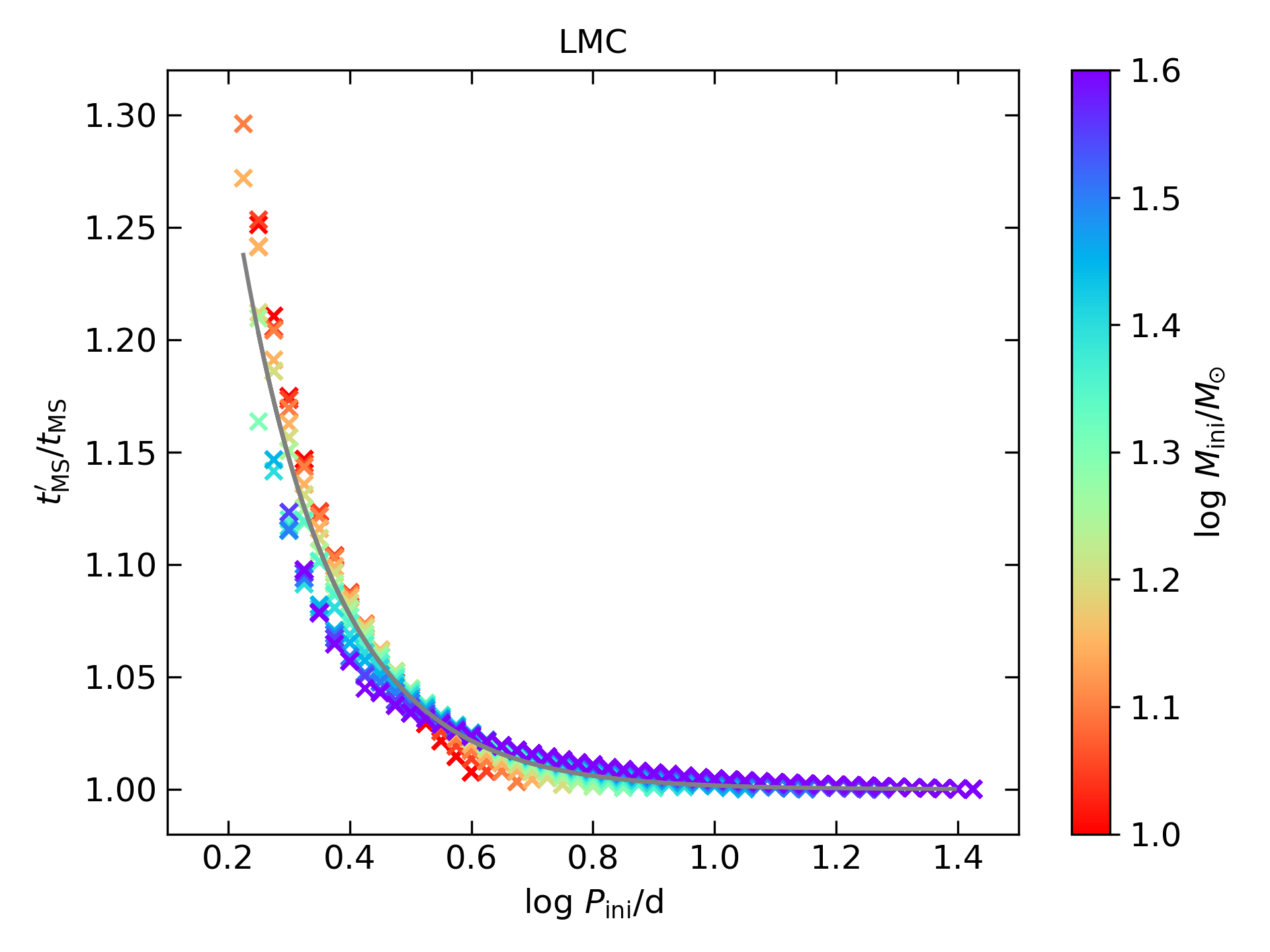}
    \includegraphics[width=0.5\hsize]{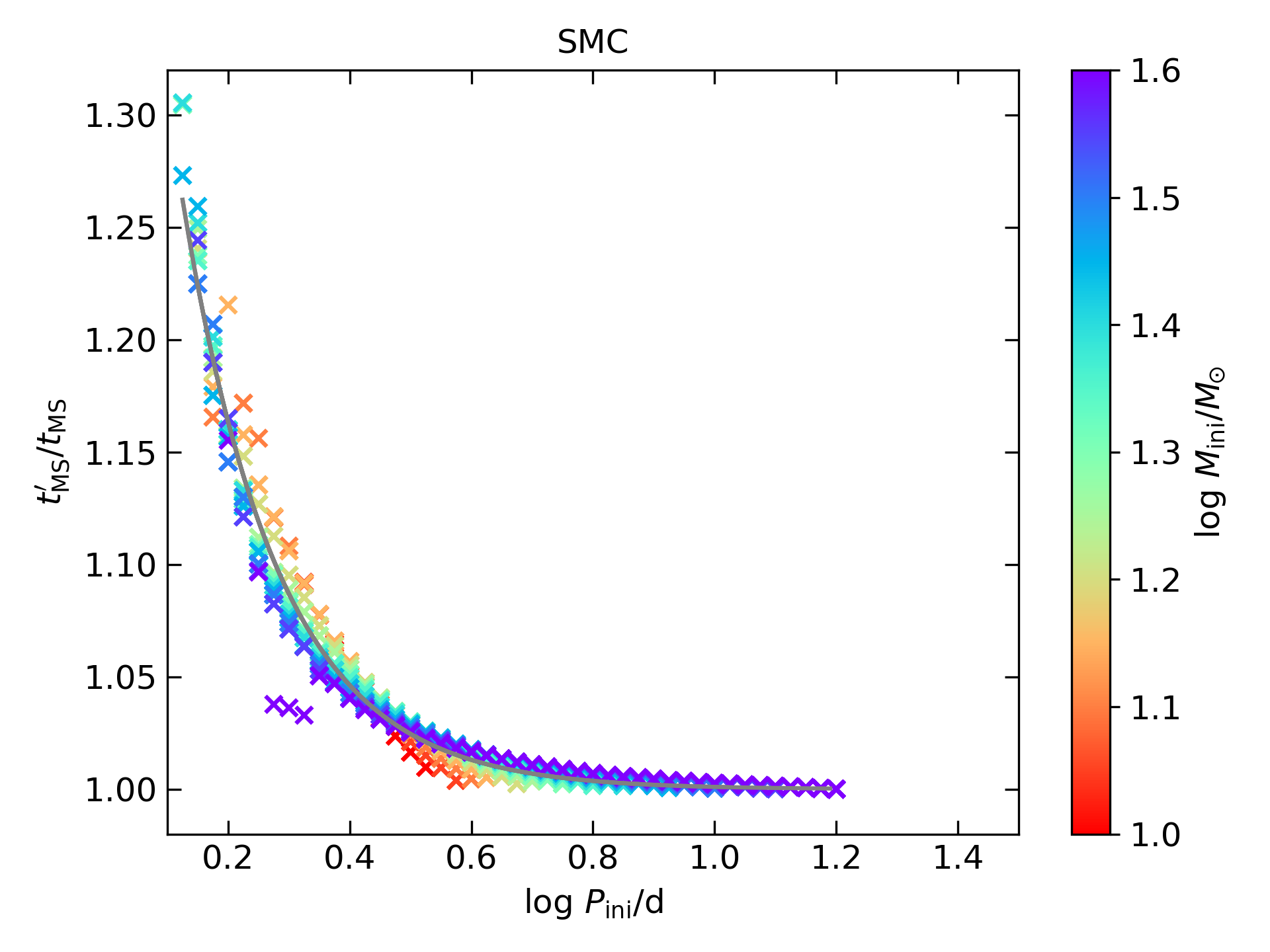}
    \caption{Duration of Case~A mass transfer ($\Delta t_\mathrm{A}$) in logarithmic years as functions of initial orbital period $P_\mathrm{ini}$ with the initial donor mass $M_\mathrm{ini}$ colour coded (top) and as functions of the initial donor mass, where models with the same initial orbital period are indicated with the same colour (middle). The bottom panels show the ratio of the Case~A donor core hydrogen burning lifetime $t_\mathrm{MS}^\prime$ to the core hydrogen burning lifetime $t_\mathrm{MS}$ of a single star of the same initial mass, as a function of the initial orbital period and in the top plots we indicted in black its interquartile range. Each cross represents the median value of $\Delta t_\mathrm{A}$ across different initial mass ratios. In the top plot we indicted in black the first and third quartile. Grey lines indicate our best fit to the data. The panels on the left show LMC models and on the right is SMC.}
    \label{fig:dt}
\end{figure*}

Fig.~\ref{fig:dt} shows in its top and middle panels the duration of Case~A mass transfer as a function of initial donor mass and initial orbital period. We find that the duration is larger for initially closer orbits, since donor stars in close orbits fill their Roche lobe earlier and thus more of the central hydrogen burning time remains for the donor in its mass-reduced state. Furthermore, the duration of Case~A mass transfer increases weakly with increasing initial mass for initial orbital periods above about $10^{0.5\cdots0.6}\,$d and decreases for lower initial periods. This means a stronger decrease in duration of mass transfer with initial orbital period for smaller initial donor masses.
For our lowest masses ($10\msol$) with the closest orbits, we find durations for Case~A mass transfer of about $10^7$ years. Interestingly for both metallicities the Case~A duration is about $10^6$ years around initial orbital periods around $10^{0.5\cdots0.6}\,$d independently of initial donor mass. From the middle plots one can see that the differences in Case~A duration between the two metallicities are small and mainly arise from the initial masses and periods where Case~A mass transfer is stable. In particular, the upper left corner of the middle panel of Fig.~\ref{fig:dt} contains models for the LMC grid, but not for the SMC grid. It also shows through the interquartile range that the impact of the initial mass ratio is very small.

In the bottom plot of Fig.~\ref{fig:dt}, we show the ratio of the core hydrogen burning lifetime of the Case~A donor $t_\mathrm{MS}^\prime$ in units of the core hydrogen burning lifetime of a single star of the same initial mass $t_\mathrm{MS}$. We find that in this representation a strong non-linear correlation to the initial orbital period. The lifetime increases for smaller initial orbital period. This is not unexpected as systems with lower initial orbital period undergo RLO earlier, have thus a larger hydrogen fraction in the core after the fast part of the mass transfer and are less massive and therefore keep core hydrogen burning for a longer time. We find increases in lifetime of up to 30\% for the closest systems. For larger initial orbital periods, the lifetime increase becomes zero as the upper orbital period for Case~A mass transfer is reached. For the the $10\msol$-models this happens around an initial period of $10^{0.6}$\,d and for the $40\msol$-models around $10^{1.4}$\,d (LMC) and $10^{1.2}$\,d (SMC), respectively. The bottom panels show that for the same initial orbital period, the lifetime increase is larger for the larger metallicity.


We found that if we would normalise the data so that we would show the lifetime increase as a function between minimum and maximum of period where Case~A mass transfer occurs, they would not lie as neatly on one curve as shown here. Using the orbital period as the independent quantity for Fig.~\ref{fig:dm} and~\ref{fig:dt} may seem to be an arbitrary choice, but we found that only with that the data fall onto one single curve. Using the relative age of the donor at beginning of the mass transfer compared to the age of central hydrogen exhaustion of a single star of same initial mass or the central hydrogen content at beginning of the mass transfer as the independent quantity, did not yield such unique curves. For practical application, we provide these data with fits in Appendix~\ref{apx}.

\subsection{Analytic fits}\label{sec:fits}

Before we provide fits for the donor mass after Case~AB mass transfer and the duration of Case~A mass transfer, we are going to give mass dependent boundaries of initial periods in which Case~A occurs and within which our fits are valid. We find that the lower period limit $P_\mathrm{min}$ for a Case~A mass transfer which leads to donor stripping is well described by a parabola
\begin{equation}
    \log P_\mathrm{min} = a + b \cdot \left(\log M_\mathrm{ini}-c\right)^2
\end{equation}
with $(a,b,c) = (0.240\pm0.001, 0.270\pm0.134, 1.04\pm0.13)$ for the LMC and $(a,b,c) = (0.114\pm0.012, 1.72\pm0.27, 1.37\pm0.02)$ for the SMC. On the other hand the upper period limit for Case~A mass transfer $P_\mathrm{max}$, which is also the boundary towards Case~B, is also well described by a parabola
\begin{equation}
    \log P_\mathrm{max} = a + b \cdot \left(\log M_\mathrm{ini}-c\right)^2
\end{equation}
with $(a,b,c) = (0.619\pm0.022, 1.87\pm0.21, 0.957\pm0.039)$ for the LMC and $(a,b,c) = (0.535\pm0.019, 1.31\pm0.13, 0.897\pm0.040)$ for the SMC.

We find third order polynomials $f$ well fitting to describe the dependency of the donor mass $M_\mathrm{AB}$ after Case~AB mass transfer and its duration $\Delta t_\mathrm{A}$ on the initial donor mass $M_\mathrm{ini}$ and the initial orbital period $P_\mathrm{ini}$. We define $m = \log M_\mathrm{ini}/\msol$ and $p = P_\mathrm{ini}/$d. With that the polynomials can be written as  
\begin{equation}\label{eq}
\begin{aligned}
    f(m, p) =\,& a_{30} m^3 + a_{20} m^2 + a_{10} m \,+ \\
             &   a_{03} p^3 + a_{02} p^2 + a_{01} p \,+ \\
             &   a_{21} m^2 p + a_{12} m p^2 + a_{11} m p + a_{00}. \\
\end{aligned}    
\end{equation}
The coefficients $a_{ij}$ of the fit are given in Table~\ref{tab:fit} for both metallicities. The root-mean-square relative deviation between model data and fit is in all cases smaller than 3\% and the maximum relative deviation reaches about 15\% for the worst outlier. We conclude that our fit describes the data well. We indicated the fits in Fig.~\ref{fig:dm} and~\ref{fig:dt} (top and middle) with grey lines for selected values of the colour coordinate, which confirms that they match well.

\begin{table*}[h]
    \centering
    \caption{Fit coefficients $a_{ij}$ found for Eq.~\ref{eq}. The last rows show root-mean-square relative deviation $\delta_\mathrm{rms}$ and the maximum relative deviation $\delta_\mathrm{max}$ between fit and data.}
    \begin{tabular}{ccccc} \hline\hline
         & \multicolumn{2}{c}{LMC} & \multicolumn{2}{c}{SMC} \\
         & $f(m,p)=\log \Delta t_\mathrm{A}$ & $f(m,p)=M_\mathrm{AB}$ & $f(m,p)=\log \Delta t_\mathrm{A}$ & $f(m,p)=M_\mathrm{AB}$ \\ \hline
        $a_{30}$ & $0.958\pm0.187$& $ 20.5\pm 1.2$ & $0.356\pm0.212$& $ 18.8\pm 1.7$ \\ 
        $a_{20}$ & $ 1.95\pm0.67$ & $-56.1\pm 4.2$ & $ 1.83\pm0.80$ & $-53.0\pm 6.5$ \\ 
        $a_{10}$ & $-8.84\pm0.81$ & $ 53.2\pm 5.1$ & $-6.30\pm1.03$ & $ 51.1\pm 8.4$ \\ 
        $a_{03}$ & $-2.14\pm0.04$ & $ 9.83\pm0.27$ & $-3.27\pm0.06$ & $ 25.9\pm 0.5$ \\ 
        $a_{02}$ & $-18.8\pm 0.3$ & $ 8.46\pm 1.6$ & $-15.0\pm 0.3$ & $ 25.2\pm 2.3$ \\ 
        $a_{01}$ & $-6.27\pm0.32$ & $ 6.03\pm2.02$ & $-5.71\pm0.42$ & $ 10.0\pm 3.4$ \\ 
        $a_{21}$ & $-11.4\pm 0.3$ & $ 19.1\pm 1.8$ & $-7.88\pm0.29$ & $ 36.7\pm 2.4$ \\ 
        $a_{12}$ & $ 14.9\pm 0.2$ & $-24.4\pm 1.3$ & $ 13.3\pm0.23$ & $-55.7\pm 1.9$ \\ 
        $a_{11}$ & $ 19.0\pm 0.6$ & $-13.5\pm 3.8$ & $ 13.1\pm 0.7$ & $-32.3\pm 5.5$ \\ 
        $a_{00}$ & $ 12.9\pm 0.3$ & $-18.0\pm 2.1$ & $ 11.3\pm 0.5$ & $-16.8\pm 3.7$ \\ \hline 
        $\delta_\mathrm{rms}$ & $0.7\%$ & $2.0\%$ & $0.6\%$ & $2.7\%$ \\
        $\delta_\mathrm{max}$ & $ 5.6\%$ & $11.0\%$ & $ 5.3\%$ & $14.4\%$ \\ \hline
    \end{tabular}
    \label{tab:fit}
\end{table*}

Next, we consider the donor mass after Case~AB $M_\mathrm{AB}$ in units of the donor mass after Case~B $M_\mathrm{B}$. We found a power law of the form
\begin{equation}\label{eq:mdivm}
    \frac{M_\mathrm{AB}}{M_\mathrm{B}} = 1 - a \cdot \left( \frac{P_\mathrm{ini}}{\mathrm{d}} \right)^b
\end{equation}
well fitting. We find $(a,b) = (0.841\pm0.007, -1.253\pm0.007)$ for the LMC and $(a,b) = (0.657\pm0.008, -1.487\pm0.016)$ for the SMC. The root mean square relative deviation between data and fit are 3\% and 4\% for LMC and SMC. The maximum relative deviation has relative high values of 15\% and 24\%. They can be explained with the neglection of mass dependence and the wavy structure in the period dependence (consider e.g. the purple sequence in Fig.~\ref{fig:dm}, bottom). For the SMC data the deviation is so strong as they do not fall so well on a single curve due to the jump between post-Case~B mass and highest post-Case~AB mass. The fit is shown in Fig.~\ref{fig:dm} (bottom) in grey.

Finally, we give a fit for the relative increase in core hydrogen burning lifetime $t^\prime_\mathrm{MS}/t_\mathrm{MS}$ again in form of a power law
\begin{equation}\label{eq:tdivt}
    \frac{t^\prime_\mathrm{MS}}{t_\mathrm{MS}} = 1 + a \cdot \left( \frac{P_\mathrm{ini}}{\mathrm{d}} \right)^b.
\end{equation}
We found $(a,b) = (1.003\pm0.010, -2.779\pm0.011)$ for the LMC and $(a,b) = (0.577\pm0.005, -2.741\pm0.015)$ for the SMC best fitting. The root mean square relative deviation between data and fit are 0.6\% and 0.7\% and the maximum relative deviation has values of 5\% and 7\% for LMC and SMC. The latter is impacted by the three outliers around $P_\mathrm{ini} \approx 10^{0.3}\days$. A visual inspection of the fit in Fig.~\ref{fig:dt} (bottom, grey line) shows that it does not trace the mass dependence perfectly, i.e. that donors with lower initial mass reach unity at lower orbital periods, but such small deviations can safely be disregarded.



The fits for the post-Case~AB mass and the lifetime increase are independent of initial donor mass. This suggests that our results may be applicable outside of the considered donor mass range. To test that, we compared our fits to additional detailed models. For the LMC we used models from the extensions of the LMC grid by \citet{2022A&A...667A..58P}, and for the SMC models of our grid outside of the adopted mass range. We show in Table~\ref{tab:test} the parameters of the models and compare the outcomes of Case~A. We find that the typical deviation between fit and detailed model is less then 10\%. Only the lifetime of the $5\msol$ SMC model and the post-Case~AB mass of the $70\msol$ LMC model deviate more than that. The typical deviation is comparable with the deviations within the analysed models in Sect.~\ref{sec:analysis} and thus we conclude that our fits are also applicable outside of their original mass range, at least as long the models have similar structure (i.e. a convective core and a radiative envelope).

\begin{table*}[h]
    \centering
    \caption{Test of our fits against models outside the used mass range ($10\msol$ to $40\msol$). Columns~7 and~8 give the post-Case~AB mass and the hydrogen burning lifetime from the detailed models, while columns~9 and~10 show the results of our fits (eq.~\ref{eq:mdivm} and~\ref{eq:tdivt}) calculated from columns~1 to~6.}
    \begin{tabular}{cccccccccc} \hline\hline
        $Z$ & $M_\mathrm{ini}/\msol$ & $q_\mathrm{ini}$ & $P_\mathrm{ini}/\mathrm{d}$ & $M_\mathrm{B}/\msol$ & $t_\mathrm{MS}/\mathrm{a}$ & $M_\mathrm{AB}/\msol$ & $t^\prime_\mathrm{MS}/\mathrm{a}$ & $M_\mathrm{AB}/\msol$ & $t^\prime_\mathrm{MS}/\mathrm{a}$ \\
        \multicolumn{6}{c}{} & \multicolumn{2}{c}{(detailed models)} & \multicolumn{2}{c}{(our fit)} \\ \hline
        LMC & 50 & 0.7  & $10^{0.45}$ & 28 & $4.5\cdot10^6$ & 22 & $4.6\cdot10^6$ & 22 & $4.8\cdot10^6$ \\
        LMC & 70 & 0.65 & $10^{0.55}$ & 45 & $3.7\cdot10^6$ & 32 & $3.8\cdot10^6$ & 37 & $3.8\cdot10^6$ \\
        SMC & 5.0 & 0.85& $10^{0.3}$  & 1.3& $1.0\cdot10^8$ & 1.0& $1.8\cdot10^8$ & 1.0& $1.1\cdot10^8$ \\
        SMC & 6.3 & 0.8 & $10^{0.125}$& 1.8& $6.1\cdot10^7$ & 0.9& $8.2\cdot10^7$ & 1.0& $7.8\cdot10^7$ \\
        SMC & 6.3 & 0.8 & $10^{0.425}$& 1.8& $6.1\cdot10^7$ & 1.5& $6.2\cdot10^7$ & 1.5& $6.3\cdot10^7$ \\
        SMC & 50 & 0.7  & $10^{0.55}$ & 29 & $4.5\cdot10^6$ & 26 & $4.6\cdot10^6$ & 26 & $4.6\cdot10^6$ \\
        SMC & 50 & 0.7  & $10^{0.7}$  & 29 & $4.5\cdot10^6$ & 27 & $4.5\cdot10^6$ & 27 & $4.5\cdot10^6$ \\
        SMC & 80 & 0.7  & $10^{1}$    & 56 & $3.5\cdot10^6$ & 52 & $3.5\cdot10^6$ & 55 & $3.5\cdot10^6$ \\
        \hline
    \end{tabular}
    \label{tab:test}
\end{table*}

\section{Discussion}\label{sec:discuss}

\subsection{Impact of the initial mass ratio and the accretion efficiency}\label{sec:qeps}

We have see in Sect.~\ref{sec:analysis}, that the post-Case~AB mass and the duration of Case~A mass transfer are nearly independent of the initial mass ratio (Fig.~\ref{fig:dm} top panels and Fig.~\ref{fig:dt} middle panels). This result is further reinforced by the small deviation between data and fits in Sect.~\ref{sec:fits}, since the fits do not consider the initial mass ratio and are still very good. We can explain the insensibility of our results to the initial mass ratio by considering each of the three phases of the RLO individually. The onset of interaction at a fixed orbital period happens more or less at the same donor radius, and thus donor age, nearly independent of mass ratio \citep[about a factor of 2 in orbital period over the relevant regime, ][]{1983ApJ...268..368E,2023arXiv231101865M}. The end of the first fast phase of mass transfer is determined by the interplay between the radius evolution of the donor star and the evolution of its Roche radius. The former is a question of stellar physics and the latter mainly depends on the orbital period and is again only a weak function of mass ratio. This is in opposition to \citet{1968ZA.....68..107G}, who predicted larger mass loss from the donor for smaller initial mass ratios (their fig.~6 in particular). The slow Case~A phase is determined by the nuclear evolution of the donor star, on which the accretor star has no impact. Finally, in Case~AB mass transfer, which is very similar to Case~B, the donor loses mass until its helium core ignites on which the companion has also no impact. All this is in agreement with fig.~F.3 to~F.5 of \citet{2022A&A...659A..98S}.

It may appear that our model grid has the shortcoming that we lose generality by assuming a certain mass transfer evolution during RLO. As the secondary star accretes matter until it reaches critical rotation, tidal forces cause a wide range of mass transfer efficiencies \citep[][fig.~F.2]{2022A&A...659A..98S}. The mass transfer efficiency controls the orbital evolution of the binary through the scheme of isotropic re-emission \citep{1997A&A...327..620S} and thus the size of the donor's Roche lobe. Thus one wonder whether our results are only valid for these assumptions. It turns out however, that the accretion efficiency only has a limited effect on the outcome of Case~A mass transfer. Consider the $40\msol$ models of the LMC grid. They show a clear structure in accretion efficiency \citep[][fig.~F.2]{2022A&A...659A..98S}. When we consider an initial period of $10^{0.6}\days$, we find for an initial mass ratio of 0.9 an overall accretion efficiency of about 80\%, and for an initial mass ratio of 0.55 an accretion efficiency of 30\%. Yet in both cases the donor mass after the mass transfer about $18\msol$ ($18.6\msol$ for the first and $17.5\msol$ for the latter), which is a strong indication that the accretion efficiency is a subdominant factor. In fact, it turns out, that in Fig.~\ref{fig:dm} (top) the slightly larger interquartile range due to different initial mass ratios in the mid-period regime is caused by the transition from high to low accretion efficiency. (Compare the periods with a larger interquartile range to fig.~F.2 of \citet{2022A&A...659A..98S}.) This should also cause the light wiggle in the model data in e.g. Fig.~\ref{fig:dm} around a period of about $10^{0.5}\days$. Hence we can quantify the impact of varying accretion efficiency and argue that its effect is small for the whole of both model grids.

\subsection{Metallicity dependence}

In Sect.~\ref{tab:fit}, we have found that the exponent $b$ of the power-law describing the ratio of post-Case~A mass to post-Case~B mass is smaller for the smaller metallicity. This means that this ratio increases slower with initial orbital periods for the smaller metallicity. However, the smaller offset $a$ in the power-law causes the curve of the smaller metallicity to lie on top of the other in the considered period range. Therefore we conclude that the donor mass after Case~A compared with the mass after Case~B is larger for smaller metallicities, given the same initial orbital period, and thus models with Galactic metallicity might be even lighter after Case~A mass transfer. This can be qualitatively understood by the fact that donor radii at zero-age main-sequence are larger and at higher metallicities and will fill their Roche lobe earlyer during hydrogen burning. Thus they deviate stronger from Case~B evolution.

For the increase in the core hydrogen burning lifetime, we found for both metallicities about same exponent in the power-law fit, namely about $-2.8$, and that the offset $a$ is larger for the larger metallicity. Therefore the lifetime increase for the same initial orbital period is larger for the larger metallicity. This fits with the smaller post-Case~AB mass for the larger metallicity. Again, we can extrapolate to Galactic metallicity and expect an even stronger lifetime increase.

\subsection{Other work}

The models we analysed in this work were already compared to observations. \citet{2022A&A...659A..98S} analysed LMC and Milky Way Algol binaries, which are believed to be a product of Case~A mass transfer, with the LMC grid and found a good agreement. \citet{2023A&A...672A.198S} used an extension of the LMC grid by \citet{2022A&A...667A..58P} to explain so called revered Algols in the Tarantula Nebula. \citet{2020ApJ...888L..12W}, \citet{2022NatAs...6..480W} and \citet{2023A&A...670A..43W} found a good agreement of their SMC models with the morphology of Hertzsprung-Russell diagrams, in which systems in slow Case~A mass transfer contribute to an extended main-sequence turnoff and blue stragglers.

To calculate the outcome of Case~AB mass transfer, several schemes have been adopted in the literature. The BSE-code \citep{hurley02} and its derivatives \texttt{binary\_c} \citep{2004MNRAS.350..407I,2006A&A...460..565I,2009A&A...508.1359I,2015ApJ...805...20S} and COMPAS \citep{2017NatCo...814906S,2022ApJS..258...34R} determine the post-Case~AB donor mass by removing the minimal mass necessary to keep the donor within its Roche-lobe. This process seems to be based on single star models which neglects the more complex structure (in particular the large helium enriched layer) of a model undergoing Case~A mass transfer. \citet{2023MNRAS.524..245R} proposed a simple approximation for the post-Case~AB donor mass by multiplying the post-Case~B donor mass by the relative age of the donor at the beginning of mass transfer. We show in Fig.~\ref{fig:tau}, that their method can lead to mismatches of up to 60\%. \citet{1968ZA.....68..107G} follows a more sophisticated approached using generalised main-sequences (stationary models with a particular total mass, central helium abundance and core mass). However, their figure~6 shows a clear dependency of the donor mass after fast Case~A on the initial mass ratio, which is in contradiction with our results from detailed models. The \textsc{ComBinE}-code of \citet{2018MNRAS.481.1908K} use the same approach as for Case~B to evaluate Case~A. They assume that the donor is reduced to its helium core mass. We show in this work that this assumption can be inaccurate to varying degrees, as for core hydrogen burning models, no helium core can be defined and therefore rely on the helium mass in the convective core. For the duration of Case~A they use the thermal timescale, which strongly underestimates its real duration.



\section{Conclusions and Outlook}\label{sec:concl}

In this study we analysed large grids of detailed massive binary evolution models to provide simple recipes for the donor mass after Case~AB mass transfer and for the duration of Case~A mass transfer. We found that these two quantities are nearly independent of the initial mass ratio of the binary. For the post-Case~AB donor mass relative to the post-Case~B donor mass and for the ratio of core hydrogen burning lifetime compared to that of a single star, we found that power laws (Eq.~\ref{eq:mdivm} and~\ref{eq:tdivt}) describe the models well. The main sequence lifetime of Case~A donors exceed that of single stars or Case~B donors of the same initial mass by a factor up to 30\% which depends on the initial orbital period, but which is nearly insensitive to the initial donor mass (Fig.~\ref{fig:dt}, bottom). The donor mass after Case~AB can be up to 50\% smaller than after a corresponding Case~B mass transfer (Fig.~\ref{fig:dm}, bottom). We predict lighter donors after mass transfer and a larger lifetime increase at higher metallicities for given initial orbital periods. We found that our results are independent on the employed mass transfer efficiency and found evidence, that our results are also valid outside the considered mass range.

While the qualitative idea is already there in the literature, this work quantifies them such that the results can be implemented into rapid binary population synthesis codes. They can be used to update the predictions of gravitational wave merger rates for different classes of compact objects. This result could also important for the predicted number of double white dwarf binaries in the Milky Way that LISA can detect. In a forthcoming paper we will use our recipe in a rapid binary population synthesis of the post-mass transfer massive star population of the SMC.

\begin{acknowledgements}
 PM acknowledges support from the FWO senior postdoctoral fellowship No.~12ZY523N.
\end{acknowledgements}

\bibliographystyle{aa}
\bibliography{bib}

\appendix
\onecolumn
\section{Fits as a function of main-sequence lifetime}\label{apx}

\begin{figure*}[h]
    \includegraphics[width=0.5\hsize]{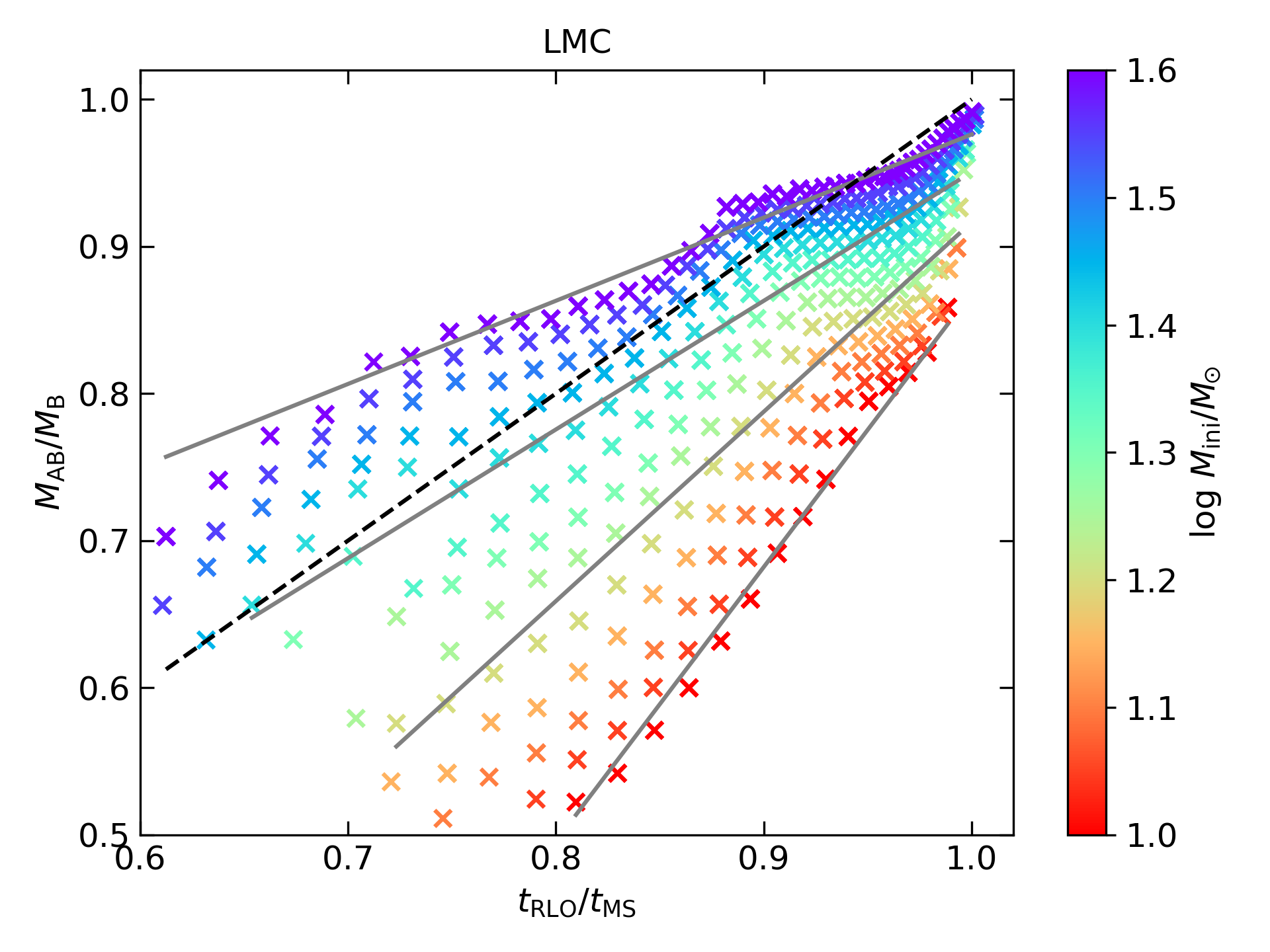}
    \includegraphics[width=0.5\hsize]{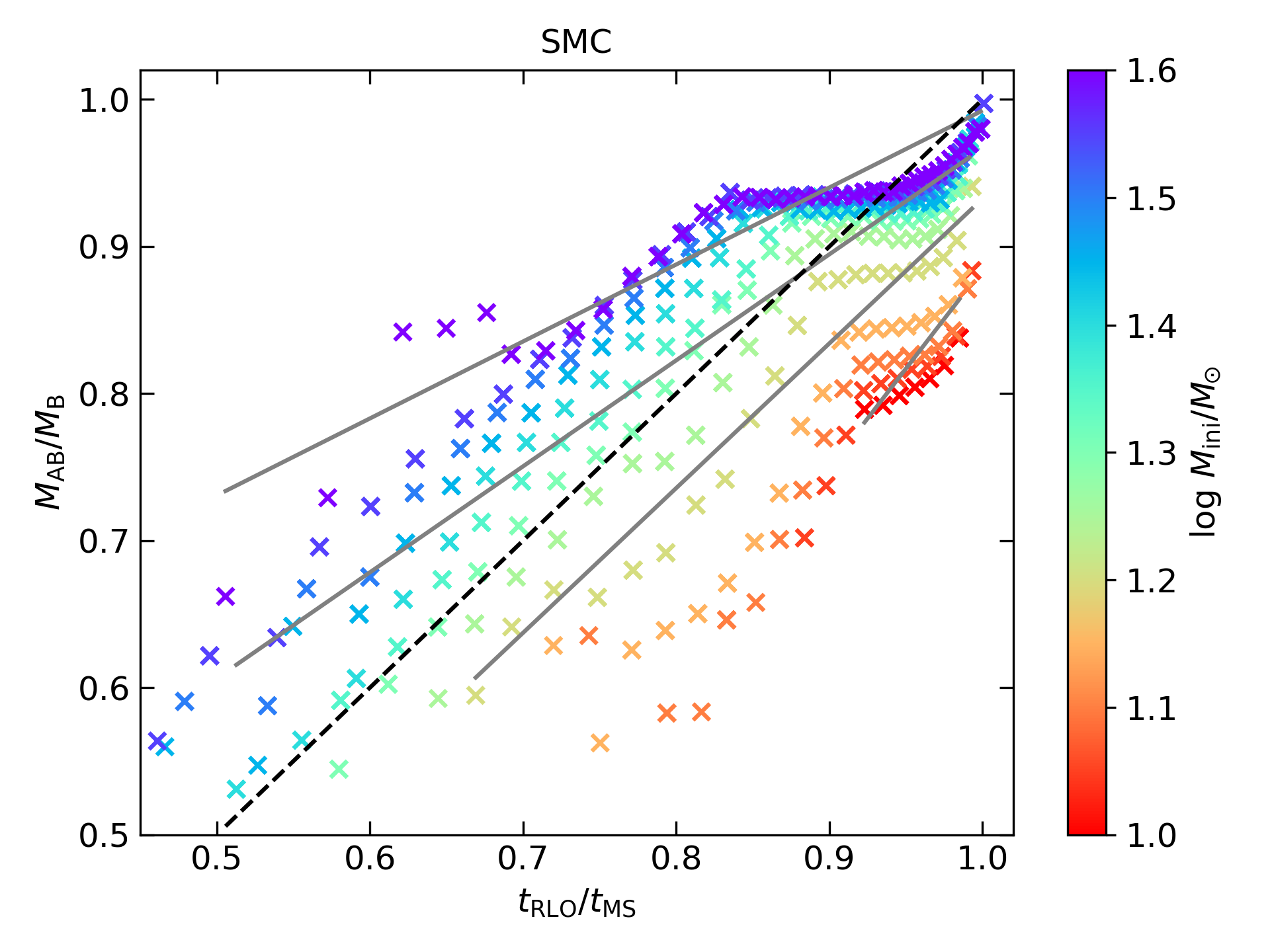}
    \includegraphics[width=0.5\hsize]{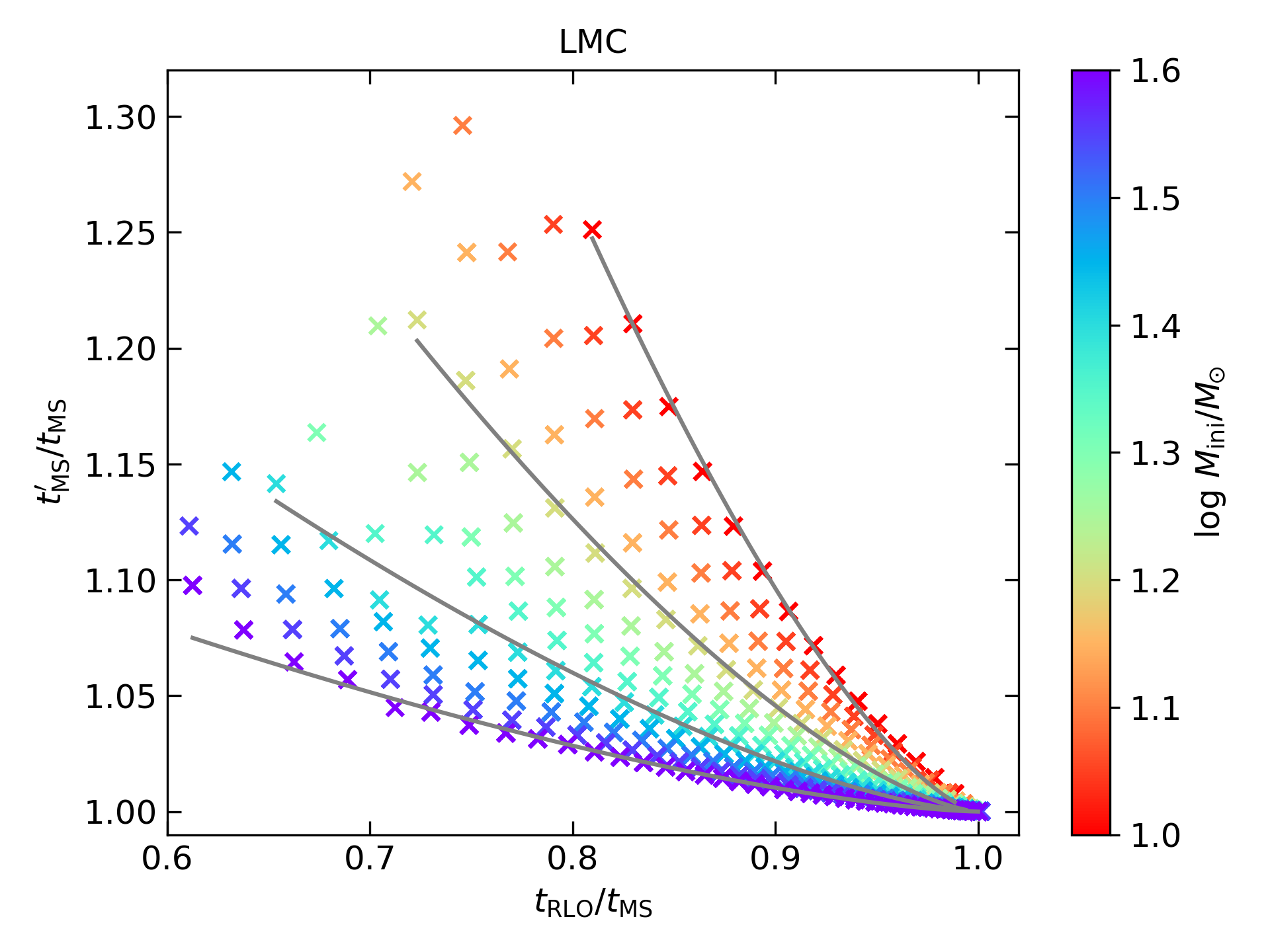}
    \includegraphics[width=0.5\hsize]{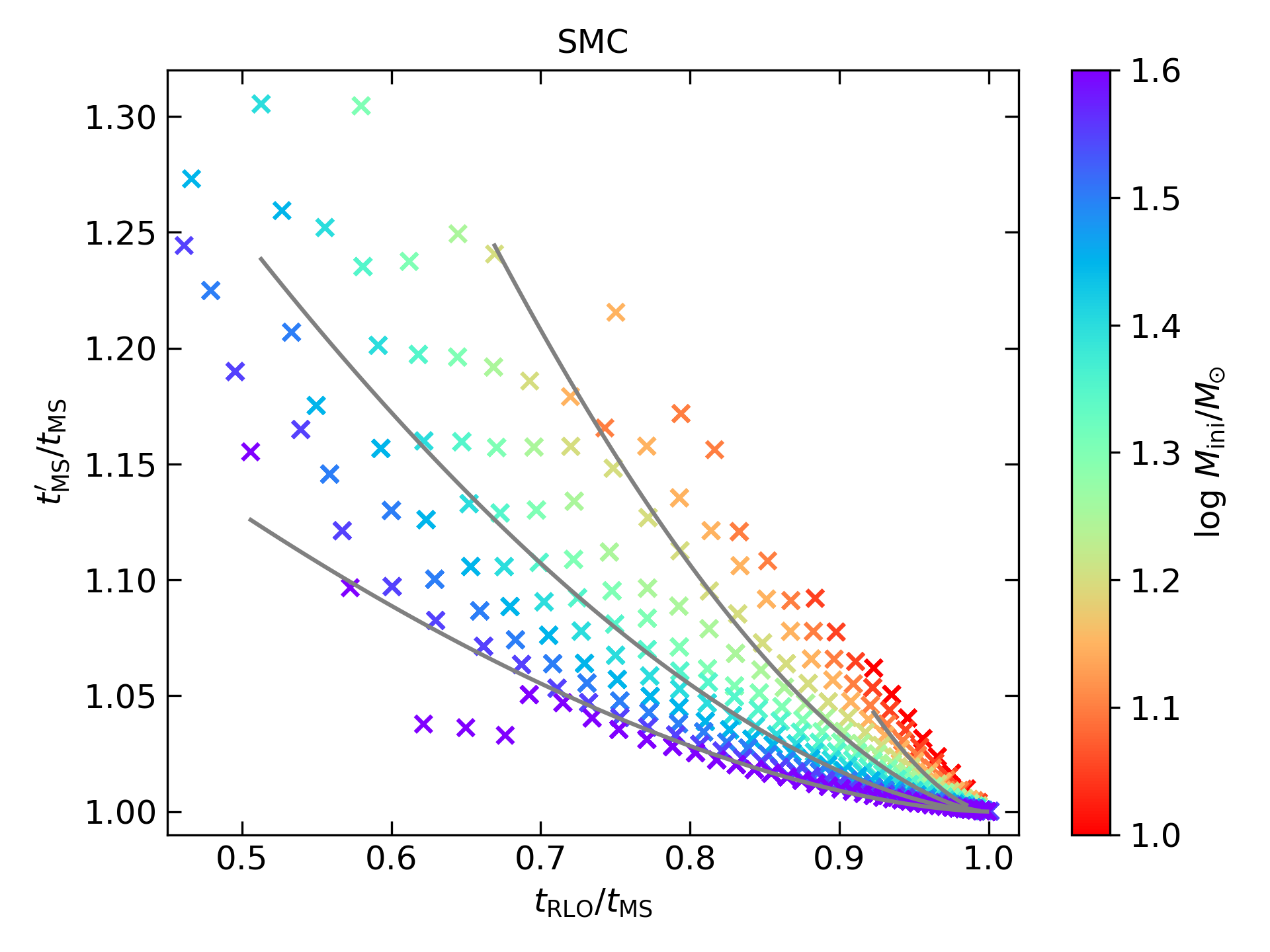}
    \caption{Same as Fig.~\ref{fig:dm} (top, also top here) and~\ref{fig:dt} (top, here bottom), but as a function of the fraction in donor hydrogen burning lifetime when the RLO begins. Grey lines indicate our best fit to the data and the black dashed line shows the approach of \citet{2023MNRAS.524..245R}, i.e. $M_\mathrm{AB} = M_\mathrm{B} t_\mathrm{RLO}/t_\mathrm{MS}$. The panels on the left show LMC models, and on the right SMC show models.}
    \label{fig:tau}
\end{figure*}

For the application of our results, it may be more practical to use the ratio of the time $t_\mathrm{RLO}$ when the RLO begins to the main-sequence lifetime $t_\mathrm{MS}$ of the donor as the independent quantity instead of the initial orbital period. We show that in Fig.~\ref{fig:tau}. The ratio of the post-Case~AB mass to the post-Case~B mass increases and the core hydrogen burning lifetime decreases with $t_\mathrm{RLO}/t_\mathrm{MS}$. In this representation the two quantities become mass dependent again. For the mass after Case~AB, we find a rational function of the form
\begin{equation}
    \frac{M_\mathrm{AB}}{M_\mathrm{B}} = a + \frac{b}{\log M_\mathrm{ini}} + \frac{ct_\mathrm{RLO}}{t_\mathrm{MS}} + \frac{d t_\mathrm{RLO}}{t_\mathrm{MS} \log M_\mathrm{ini}}
\end{equation}
well fitting. We find $(a,b,c,d) = (2.76\pm0.03,-3.75\pm0.05,-1.60\pm0.04,3.47\pm0.05)$ for the LMC and $(a,b,c,d) = (2.03\pm0.07,-2.50\pm0.10,-0.86\pm0.08,2.21\pm0.11)$ for the SMC. The root mean square relative deviations are 2\% and 4\%, and the maximum relative deviations are 12\% and 21\%.

For the increase in core hydrogen burning lifetime, we find a power law of the form 
\begin{equation}
    \frac{t_\mathrm{MS}^\prime}{t_\mathrm{MS}} = 1 + a \cdot M_\mathrm{ini}^{-b} \cdot \left(1-\frac{t_\mathrm{RLO}}{t_\mathrm{MS}}\right)^c
\end{equation}
well fitting. We find $(a,b,c) = (116.8\pm1.7,1.618\pm0.004,1.465\pm0.003)$ for the LMC and $(a,b,c) = (80.5\pm4.7,1.438\pm0.016,1.649\pm0.012)$ for the SMC. The root mean square relative deviations are 0.3\% and 0.9\%, and the maximum relative deviations are 4\% and 6\%.

\end{document}